\documentclass[]{article}
\usepackage{graphicx}
\usepackage{color}
\usepackage{amsfonts}
\usepackage{amssymb}
\def\ligne#1{\hbox to\hsize{#1}}
\def\leurre{\noindent\leftskip0pt\small\baselineskip 10pt}
\newtheorem{thm}{\textrm{\sc Theorem}}

\newtheorem{fig}{\textrm{Figure}}
\newtheorem{tab}{\textrm{Table}}
\def\boxempty{\hbox{\vbox{\hsize=7pt\offinterlineskip
\ligne{
\vrule height 7pt depth 0pt width 0.6pt
\vbox to 7pt{\hsize=5.8pt
\hrule height 0pt depth 0.6pt width 5.8pt
\vfill
\hrule height 0.6pt depth 0pt width 5.8pt
}\hskip-0.5pt
\vrule height 7pt depth 0pt width 0.6pt
}}
}}
\def\trep{\hrule height 1pt depth 1pt width \hsize}
\def\trfn{\hrule height 0.5pt depth 0.5pt width \hsize}
\newcounter{laform}
\author{Maurice {\sc Margenstern}}
\title{A weakly universal cellular automaton on the tessellation $\{8,3\}$.}
\begin{document}
\maketitle

\begin{abstract}
In this paper, we construct a weakly universal cellular automaton on the tessellation
$\{8,3\}$ which is not rotation invariant but which is truly planar. 
\end{abstract}

\section{Introduction}

   This paper is basically an improvement of paper~\cite{mmarXiv1605b} where I proved the
same result in the tessellation $\{9,3\}$. The reason of this improvement lies in
the relatively small number of rules for paper~\cite{mmarXiv1605b} and the fact, noticed
in that paper, that several rules where uselessly duplicated. Also, as it is usual in 
this process of reducing the possibilities of the automaton, here its neighbourhood,
it is needed to change something in the previous scenario of the simulation.
This time, I simplified a structure involved in the simulation by replacing it
by the combination of two other existing structures. 
The morale of this result compared to that of~\cite{mmJCA2016} is that relaxing the
rotation invariance allows us to significantly reduce the number of neighbours: from~11
in~\cite{mmJCA2016} to~8 in the present paper. As in~\cite{mmarXiv1605b},
I use the new system of coordinates introduced in~\cite{mmarXiv1605a} for the
tilings $\{p,3\}$ and $\{p$$-$$2,4\}$. 

   In this paper, the same model as in~\cite{mmJCA2016} and~\cite{mmarXiv1512} and the
other quoted papers is used. 

In Section~\ref{scenar}, I just indicate how the model is tuned a bit in order to
replace the structure above evoked by new ones. In Section~\ref{rules}, we give the
rules of the automaton, insisting in the way we defined these rules in a context where
rotation invariance is no more required, which allows us to prove the following result:

\begin{thm}\label{letheo}
There is a weakly universal cellular automaton on the tessellation $\{8,3\}$
which is 
truly planar and which has two states.
\end{thm}

Presently, we turn to the proof of this result.

\section{The scenario of the simulation}
\label{scenar}

   I reproduced the model used for these simulations in each paper. This time, to
spare space and time, I just mention where the reader can find a description of the
model and what is changed with respect to the previous simulation for the
tessellation~$\{9,3\}$.

    We sketchily remember that we simulate a register machine by a railway circuit.
Such circuit assembles infinitely many portions of straight lines, quarters of circles 
and switches. There are three kinds of switches, see~\cite{stewart,mmbook3}, for
a description of the circuit and of its working. In~\cite{mmbook3} the simulation
is thoroughly described and it is adapted to the hyperbolic context.

    As in previous papers, the flip-flop and the memory switch are decomposed into
simpler ingredients which we call sensors and control devices. This reinforces the
importance of the tracks as their role for conveying key information is more and more 
decisive. Here too, tracks are blank cells marked by appropriate black cells we call
\textbf{milestones}. We carefully study this point in Sub-section~\ref{tracks}.  Later, 
in Sub-section~\ref{struct}, we adapt the configurations described in~\cite{mmarXiv1605b}
to the tessellation $\{8,3\}$.

\subsection{The tracks}\label{tracks}

In this implementation, the tracks are represented in a 
way which is a bit similar to that of~\cite{mmarXiv1605b}. The present implementation
is given by Figure~\ref{elemtrack}. 

\vskip 10pt
\vtop{
\ligne{\hfill
\includegraphics[scale=1]{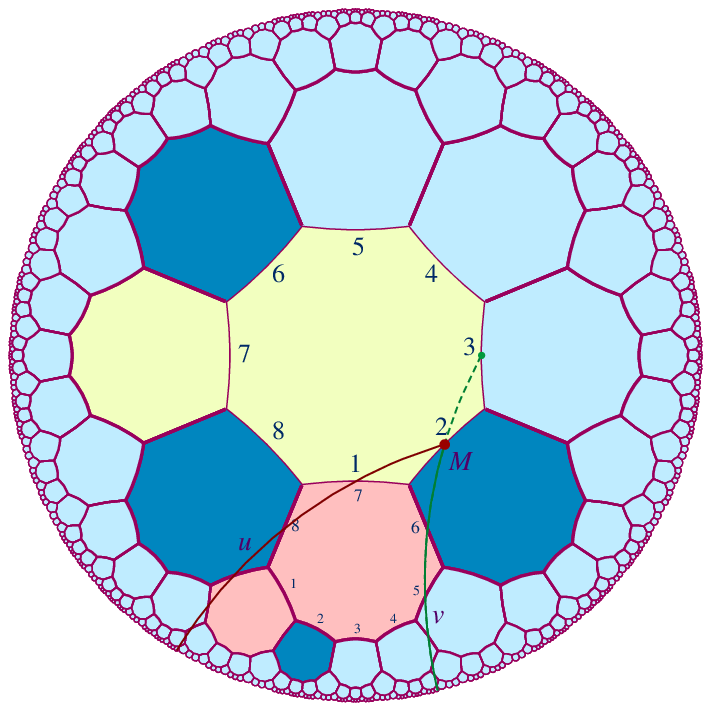}
\hfill}
\vspace{-10pt}
\ligne{\hfill
\vtop{\leftskip 0pt\parindent 0pt\hsize=160pt
\begin{fig}\label{elemtrack}
\leurre
Element of the tracks.
\end{fig}
}
\hfill}
}
\vskip 5pt
Here, we explicitly indicate the numbering of
the sides in a cell which will be systematically used through the paper. We fix a 
side which will be, by definition, side~1 in the considered cell. Then, all the other 
sides are numbered starting from this one and growing one by one while counter-clockwise 
turning around the cell. Note that, in our setting, the same side, which is shared by 
two cells, can receive two different numbers in the cells which share it. An example of 
this situation is given in Figure~\ref{elemtrack}: in the central cell we denote
by~0(0), side~1 is side~7 
in the neighbour of the central cell sharing this side. In Sub-section~\ref{srtrack}, 
we go back to the construction of the tracks starting from the elements indicated
in Figure~\ref{elemtrack}. Note that Figure~\ref{elemtrack} shows us two rays starting
from~$M$, the mid-point of the side~2 of the central cell. These rays allow us to
introduce the numbering of the tiles based on~\cite{mmarXiv1605a}. It will be used
in the figures illustrating the paper.

The rays delimit what we call a \textbf{sector}. The rays are defined as follows.
The ray~$u$ starts from the mid-point~$M$ of the side~2 of~0(0) and it
passes through the mid-point of its side~1. The ray also passes through the mid-point
of the side~8 of the neighbour of~0(0) which is seen through its side~1 and which we
denote 1(1). It also passes through the mid-point of the side~8 of~1(1).
The ray~$v$, also issued from~$M$, cuts the sides~6 and~5 of~1(1) at their mid-points. 
Its support also passes through the mid-point of the side~3 of~0(0).

In the figures of the paper, the central cell
is the tile whose centre is the centre of the circle in which the figure is inscribed.
The central cell is numbered by~0, denoted by~0(0). We number the sides of the tile
as indicated in Figure~\ref{elemtrack}. For $i\in\{1..8\}$, the cell which
shares the side~$i$ with the central cell is called \textbf{neighbour}~$i$ and it
is denoted by~$1(i)$. The rotation around~0(0) allows us to attach a sector to each
tile~$1(i)$. Number~1 in this notation is the number given to the root of the
tree attached to the sector defined from this tile, see~\cite{mmbook2,mmarXiv1605a}.
Here, that definition is adapted to the case of the 
tessellation~$\{8,3\}$. We invite the
reader to follow the present explanation on Figure~1. 
Consider the sector defined by the rays~$u$ and~$v$. The neighbours of the 
cell~1(1) sharing its sides~$j$, $j\in\{1..4\}$ are numbered~$j$+1 and are denoted
by $j$+1(1). We say that the cell~1(1) is a $W$-cell and its sons are defined
by the rule \hbox{\tt W $\rightarrow$ BWWW}, which means that 2(1) is a $B$-cell.
This means that the sons of~2(1) are defined by the rule \hbox{\tt B $\rightarrow$ BWW},
where the $B$-son has two consecutive sides crossed by~$u$ in their mid-points. 
These sons of~1(1) constitute the level~1 of the tree.
The sons of~2(1), starting from its $B$-son are numbered 6, 7 and~8 denoted by
6(1), 7(1) and 8(1) respectively. By induction, the level~$n$+1 of the tree
are the sons of the cells which lie on the level~$n$. The cells are numbered from
the level~0, the root, level by level and, on each level from left to right, {\it i.e.}
starting from the ray~$u$ until the ray~$v$. What we have seen on the numbering of the
sons of~2(1) is enough to see how the process operates on the cells. From now on, we
use this numbering of the cells in the figures of the paper.

As can be seen in Figures of Section~\ref{rules}, the locomotive is implemented as
a single black cell: it has the same colour as the milestones of the tracks. Only the 
position of the locomotive with respect to the milestones allows us to distinguish it 
from the milestones. As clear from the next sub-section, we know that besides this
\textbf{simple locomotive}, the locomotive also occurs as a \textbf{double one} in some
portions of the circuit. In a double locomotive, we call its first, second cell the 
\textbf{front}, \textbf{rear} respectively of the locomotive. In a simple locomotive,
the front only is present.

The circuit also makes use of signals which are implemented in the form of a simple 
locomotive. So that at some point, it may happen that we have three
simple locomotives travelling on the circuit: the locomotive and two auxiliary signals 
involved in the working of some switch. For aesthetic reasons, the black colour which 
is opposed to the blank is dark blue in the figures.

\subsection{The structures of the simulation}
\label{struct}

    The crossings of~\cite{stewart} are present in many ones of my papers. Starting 
from~\cite{mmarXiv1202}, I replaced the crossing by round-abouts, a road traffic
structure, in my simulations in the hyperbolic plane. At a round-about where two 
roads are crossing, if you want to keep the direction arriving at the round-about, 
you need to leave the round-about at the second road. I refer the reader 
to~\cite{mmbook3} for references. The structure is a complex one, which requires
a fixed switch, a doubler and a selector. Other structures are used to simulate the
switches used in~\cite{stewart,mmbook3}: the fork, the controller and the sensor.
In this section, we present the implementation of these structures which are those
of~\cite{mmarXiv1605b} adapted to the present tessellation with one exception: the
doubler, which is here different from that of that paper.

\subsubsection{The fixed switch}
\label{ssfx}

    As the tracks are one-way and as an active fixed switch always sends the locomotive 
in the same direction, no track is needed for the other direction: there is no active 
fixed switch. Now, passive fixed switches are still needed as just seen in the previous 
paragraph.  

Figure~\ref{stab_fx} illustrates the passive fixed switch when there is no 
locomotive around: we say that such a configuration is \textbf{idle}. We shall
again use this term in the similar situation for the other structures. 
We can see that it consists of elements of the tracks which are 
simply assembled in the appropriate way in order to drive the locomotive to the bottom 
direction in the picture, whatever upper side the locomotive arrived at the switch. 
The path followed by the locomotive to the switch is in yellow until the central
cell which is also yellow. The path from the left-hand side consists, in this order
of the cells 13(8), 4(8), 5(8), 2(1) and 1(1). From the right-hand side, it consists 
of the cells 16(6), 4(6), 5(6), 2(7) and~1(7). Of course, 1(1) and~1(7) are neighbours
of~0(0). The path followed by the locomotive when it leaves the cell is in pink.
It consists of the following cells in this order: 1(4), 2(4), 5(3), 4(3)and 16(3).
Note that the cell~0(0) in Figure~\ref{stab_fx} has five black neighbours:
the cells 1(2), 1(3), 1(5), 1(6) and~1(8). Note that 1(6) and 1(8) are also
milestones for the cell 1(7), that 1(3) and~1(5) are milestones for 1(4) and
that 1(8) and~1(2) are milestones for 1(1).

From our description of the working of the round-about, a passive fixed switch must be 
crossed by a double locomotive as well as a simple one. Later, in 
Subsection~\ref{srfx}, we shall check that the structure illustrated by
Figure~\ref{stab_fx} allows those crossings. 

\vtop{
\ligne{\hfill
\includegraphics[scale=1.1]{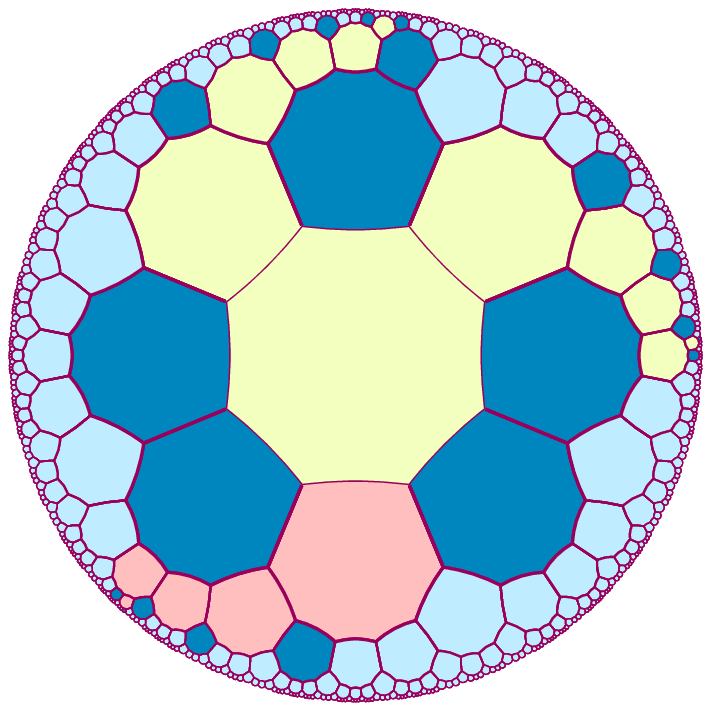}
\hfill}
\vspace{-10pt}
\ligne{\hfill
\vtop{\leftskip 0pt\parindent 0pt\hsize=270pt
\begin{fig}\label{stab_fx}
Idle configuration of the passive fixed switch.
\end{fig}
}
\hfill}
\vskip 10pt
}

\subsubsection{The doubler and the fork}
\label{ssdblfrk}

    The fork is the structure illustrated by the left-hand side picture of
 Figure~\ref{stab_dblfrk}. Note that its structure is very different 
from that of the tracks or of the fixed switch. The central cell~0(0) is 
black and two paths start from~1(1), each one on one side of the central cell
with respect to its axis crossing its sides~1 and~5, and traversing a quarter of the
cells around~0(0). The cell~1(1) is yellow in the figure. The left-hand side track is 
green, consisting of the following cells, in this order: 1(2), 1(3), 3(3), 9(3) 
and~35(3). The right-hand side track is pink. It consists of the cells: 1(8), 1(7), 5(7),
6(8) and 21(8). The locomotive, a simple one, arrives through the yellow path: 
21(2), 6(2), 5(1), and 1(1). From 1(1), two simple locomotives appear: one in 1(2), 
the other in 1(8). The locomotive in 1(2)
goes along the light green path while the one in 1(8) goes along the pink path.

   The doubler is illustrated by the right-hand side of Figure~\ref{stab_dblfrk}.
This structure is different from that of~\cite{mmarXiv1605b}. Indeed, the even number
of sides of a tile does not allow us to divide the path around 0(0) in two equal sub-paths
both excluding 1(1). The doubler is a structure which receives a simple locomotive
and yields a double one which consists of two consecutive black cells on the track.
The idea is to use two already defined structures: the fork and the fixed switch. This 
combination can make a double locomotive provided that two simple ones arrive at the
fixed switch from each side with a delay of one top of the clock. This is realized by 
the paths illustrated by the figure. The picture uses the same colours as the picture of
the fork with the same meaning. Consider the green paths. Its cells are, in this
order: 1(2), 2(3), 3(3), 4(3), 5(3), 2(4) and 3(4), which makes 7~cells. The pink path
consists of the following cells, in this order: 1(8), 1(7), 1(6), 1(5), 2(5) and 5(4).
We can see that the cells around 4(4) are exactly the neighbours of the central cell
of a fixed switch, see Figure~\ref{stab_fx}. According to this description,
the two simple locomotives created at the same time in 1(2) and 1(8) respectively
do not arrive at the same time at the cell~4(4). When the locomotive created in~1(8)
arrives at the cell~4(4), the locomotive created in 1(2) is at 3(4), so that
the two black cells in 3(4) and~4(4) constitute a double locomotive arriving to
the fixed switch from the left-hand side. We shall see that the double locomotive
does cross the switch. Accordingly the structure works as expected for a doubler.
Note the configuration of three black cells around a common vertex.

\vskip 10pt
\vtop{
\ligne{\hfill
\includegraphics[scale=0.7]{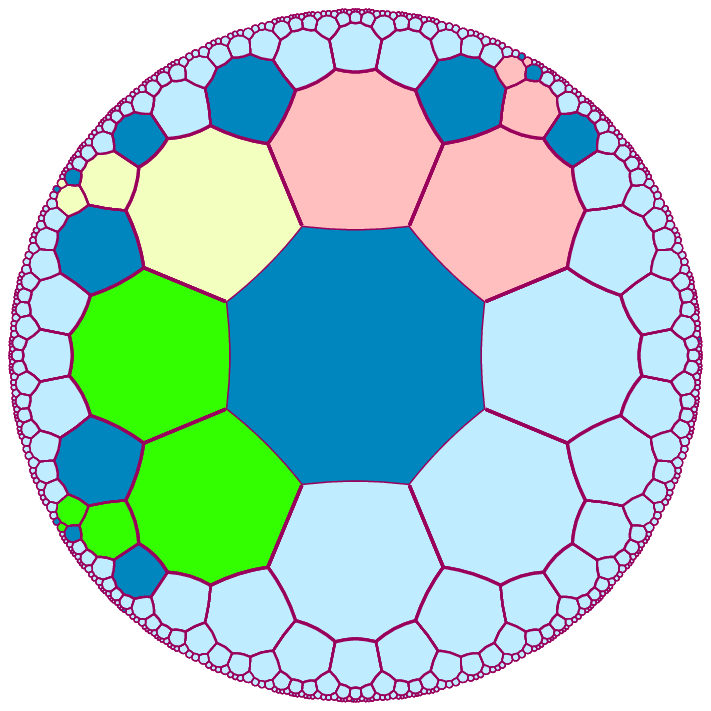}
\includegraphics[scale=0.7]{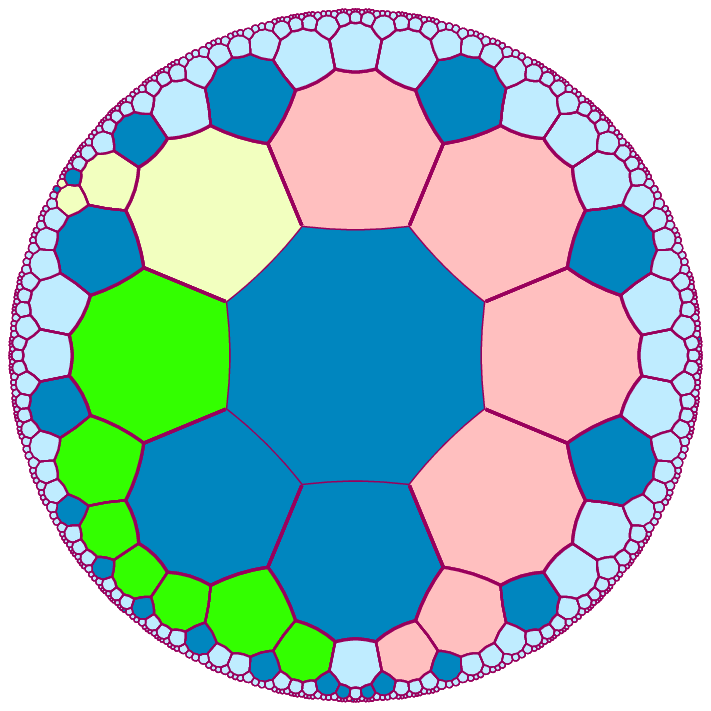}
\hfill}
\vspace{-15pt}
\ligne{\hfill
\vtop{\leftskip 0pt\parindent 0pt\hsize=230pt
\begin{fig}\label{stab_dblfrk}
\leurre
Idle configurations.
To left: the fork. To right: the doubler.
\end{fig}
}
\hfill}
}

\subsubsection{The selector}
\label{ssel}

The selector is illustrated by Figure~\ref{stab_sel}. This structure is not as symmetric
as the corresponding structure of~\cite{mmarXiv1605b}, which makes another difference
with that paper. We have a yellow track through which the locomotive arrives, simple
or double, both cases are possible. When a simple locomotive arrives, it leaves the
cell through~1(8), via the pink path which consists of the cells 1(8), 2(8), 9(8)
and 32(8). When a double locomotive arrives, a simple locomotive leaves the structure
through the green path, the cells: 1(4), 2(5), 6(5) and 24(5). Both cells 1(6) and 1(8)
can detect whether the locomotive is simple or double. They can do that when the 
front of the locomotive is in~0(0). Then, if the locomotive is double, its rear 
is in 1(6). Both cells 0(0) and~1(6) are neighbours of 1(5) and of 1(7) too.

In Subsubsection~\ref{ssel}, the rules will show that such a working will be observed.

\vtop{
\ligne{\hfill
\includegraphics[scale=1.1]{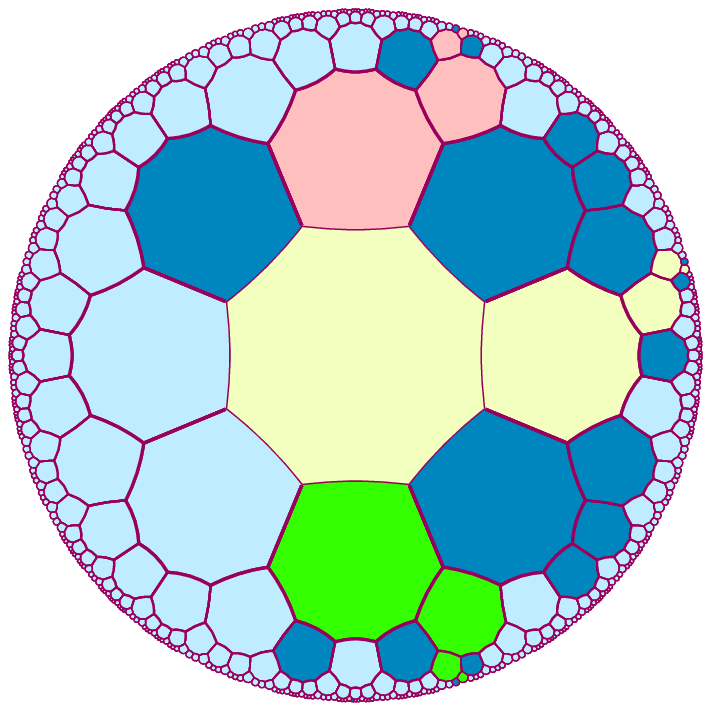}
\hfill}
\vspace{-15pt}
\ligne{\hfill
\vtop{\leftskip 0pt\parindent 0pt\hsize=230pt
\begin{fig}\label{stab_sel}
\leurre
Idle configuration of the selector. The cells~$1(7)$ and$~1(5)$ detect whether the 
locomotive is simple or double.
\end{fig}
}
\hfill}
}

\subsection{The controller and the controller-sensor}
\label{sctrlcapt}

    In this Sub-section, we look at the additional structures used for the
flip-flop and for the memory switch, see~\cite{stewart,mmbook3} for the definitions
and for the implementation in the hyperbolic plane. As explained in~\cite{mmbook3},
the flip-flop and the active memory switch are implemented by using the fixed switch,
the fork and a new structure we shall study in Subsubsection~\ref{ssctrl}: 
the \textbf{controller}. The structure is illustrated by Figure~\ref{stab_ctrl}.
For the passive memory switch, we need the fork, the fixed switch and another
new structure we shall study in Subsubsection~\ref{sscapt}: the \textbf{sensor}
illustrated by Figure~\ref{stab_capt}.

\subsubsection{The controller}\label{ssctrl}

As shown by Figure~\ref{stab_ctrl}, the controller sits on an ordinary cell
of the track. The locomotive is which runs on that track is always simple.
The track consists of the cells 20(6), 6(7), 2(7), 1(6), 0(0), 1(4), 2(5), 3(5), 10(5)
and 35(5). The cell 1(3) defines the \textbf{colour} of the controller. If it is
black, then the cell~0(0) is typically a cell of the track, so that the locomotive goes
on its way along the track, leaving the controller. If the cell 1(3) is white, then
the cell 0(0) can no more work as a cell of the track. It remains white, which means
that the locomotive is stopped at 1(6): after that, it vanishes. This corresponds
to the working of a selection in an active passage: the locomotive cannot run along
a non-selected track. Here it can for a while, but at some point, it is stopped
by the controller. Note that the occurrence of a locomotive in the structure
does not change the colour in 1(3). The change of colour in that cell is performed by
a signal which takes the view of a simple locomotive arriving through another track: 
24(4), 6(4) and 2(4), that latter cell being a neighbour of~1(3). When the
locomotive-signal arrives at 2(4), it makes the cell 1(3) change its colour: from
white to black and from black to white.

\vtop{
\vskip 0pt
\ligne{\hfill
\includegraphics[scale=1.1]{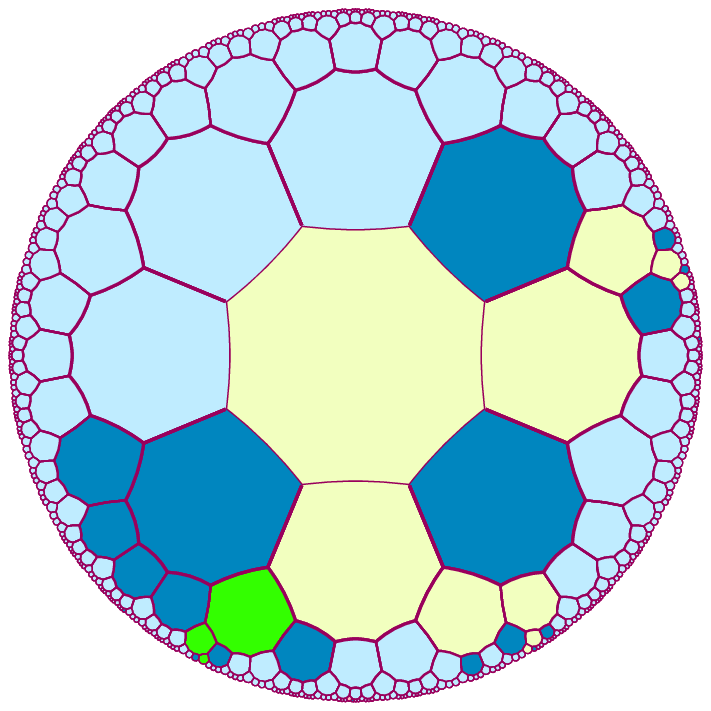}
\hfill}
\vspace{-10pt}
\ligne{\hfill
\vtop{\leftskip 0pt\parindent 0pt\hsize=320pt
\begin{fig}\label{stab_ctrl}
\leurre
Idle configuration of the controller of the flip-flop and of the active memory switch.
\end{fig}
}
\hfill}
\vskip 10pt
}

\subsubsection{The sensor}\label{sscapt}

Let us now turn to the sensor,
illustrated by Figure~\ref{stab_capt}. As suggested by its name, the sensor does not
exactly behave like the controller. When the locomotive passes on the non-selected track,
it does not stop it. It uses it as a messenger for the signal it has to send to
the active switch in order to change the selection of the tracks. 

This is illustrated by the structure of the figure. The path is the same as in 
Figure~\ref{stab_ctrl}. The cell which plays the role of a sensor is this time the
cell~1(1) whose state we call the \textbf{colour} of the sensor. Note that the 
neighbourhood of that cell in Figure~\ref{stab_capt} is the 
same, up to rotation, to the neighbourhood of the cell~1(3) in Figure~\ref{stab_ctrl}:
the green path here consists of the cells are 24(2), 6(2) and 2(2) the latter being a 
neighbour of~1(1).

Figure~\ref{stab_capt} shows a very different structure for the cell~0(0) compared
with that of Figure~\ref{stab_ctrl}. When the sensor is white, its neighbourhood
is exactly that of the cell~0(0) when the controller is black: it is an ordinary cell
of the track so that the locomotive goes on its way on the track. The difference
in both structures lies in the logic of the switches. In the case of the controller,
when the locomotive goes on its way, it is the locomotive of the circuit going to
another switch or to a round-about. In the case of the sensor, the locomotive which
goes on its way on the track becomes a signal sent to the active switch associated
to the passive switch.

\vskip 10pt
\vtop{
\ligne{\hfill
\includegraphics[scale=1.1]{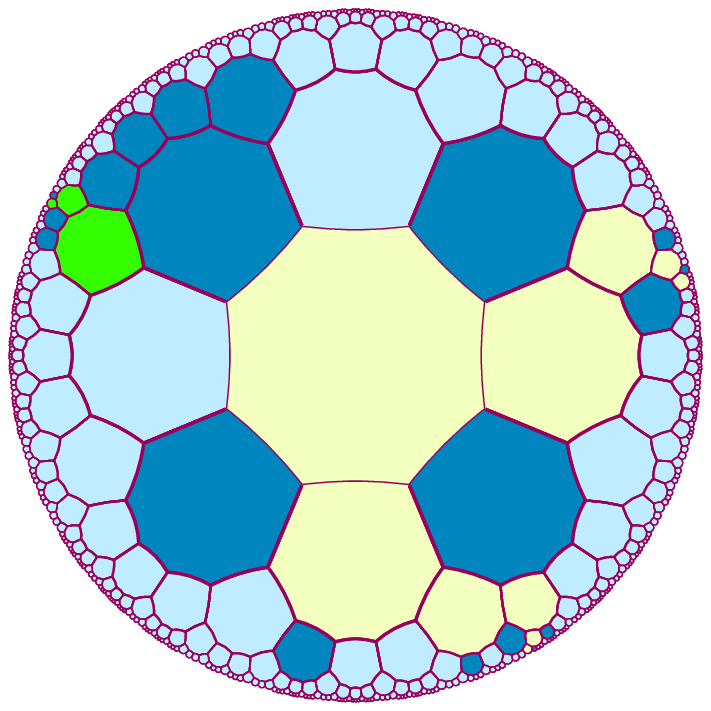}
\hfill}
\vspace{-15pt}
\ligne{\hfill
\vtop{\leftskip 0pt\parindent 0pt\hsize=300pt
\begin{fig}\label{stab_capt}
\leurre
Idle configuration of the sensor of the passive memory switch.
\end{fig}
}
\hfill}
}

We can just note that the change of colour is different in the sensor: when the
sensor is white, if a locomotive passes, it must become black: the signal is the
locomotive itself, as will be seen in the Subsection~\ref{srcapt}. When the sensor is 
black, it has to be changed if a locomotive passed through the other sensor which
then changed from white to black. The locomotive which arrived at the formerly white
sensor is sent to the still black one in order to make it change to white. The 
locomotive arrives through the green path of Figure~\ref{stab_capt}. As the configuration 
is the same around the cell~1(1) of that Figure as that around~1(3) in 
Figure~\ref{stab_ctrl}, the change from black to white is performed.

\section{Rules}
\label{rules}

    The figures of Section~\ref{scenar} help us to establish the rules. Their application
is illustrated by figures of this section which
were drawn by a computer program which checked the 
coherence of the rules. The program also wrote the PostScript files of the pictures from 
the computation of the application of the rules to the configurations of the various type 
of parts of the circuit. The computer program also established the traces of execution 
which contribute to the checking of the application of the rules.

    Let us explain the format of the rules and what is allowed
by the relaxation from rotation invariance. We remind the reader that a rule has the form
\hbox{\footnotesize\tt$\underline{\hbox{X}}$$_o$X$_1$..X$_8$$\underline{\hbox{X}}$$_n$},
where \hbox{\footnotesize\tt X$_o$} is the state of the cell~$c$,
\hbox{\footnotesize\tt X$_i$} is the \textbf{current} state of the neighbour~$i$ of~$c$ 
and \hbox{\footnotesize\tt X$_n$} is the \textbf{new} state of~$c$ applied by the
rule. As the rules no more observe the rotation invariance, we may freely 
choose which is side~1 for each cell. We take this freedom from the format of the rule 
which only requires to know which is neighbour~1. In order to restrict the number of 
rules, it is decided that as a general rule, for a cell which is an element of the track, 
side~1 is the side shared by the cell and its next neighbour on the track, so that
tracks are one way. There can be 
exceptions when the cell is in a switch or the neighbour of the central cell in a switch.
In particular, when a cell belongs to two tracks, side~1 is arbitrarily chosen among the 
two possible cases. The milestones may have their side~1 shared by an element of the 
track. 

    We have to keep in mind that there are two types of rules. Those
which keep the structure invariant when it is idle, we call this
type of rules \textbf{conservative}, and those which control the motion of the locomotive.
Those latter rules, which we call \textbf{motion rules}, are the rules applied to the
cells of  the tracks as well as their milestones and, sometimes to the cells of the 
structures which may be affected by the passage of the locomotive. Next, in each 
sub-section, we give the rules for the motion of the locomotive in the tracks,
then for the fixed switch, then for the doubler and for the fork, then for the selector,
then for the controller and, eventually, for the sensor. In each sub-section, we also 
illustrate the motion of the locomotive in the structure as well as a table giving traces
of execution for the cells of the track involved in the crossing.

\def\Rr#1{{\color{red}{#1}}}
\def\aff #1 #2 #3 #4 {\ligne{\hfill\footnotesize\tt\hbox to 13pt{\hfill#1}
\hskip 5pt$\underline{\hbox{\tt#2}}$#3$\underline{\hbox{\tt#4}}$\hfill}\vskip-4pt
}
\def\raff #1 #2 #3 #4 {\ligne{\hfill\footnotesize\tt\hbox to 13pt{\hfill{\Rr{#1}}}
\hskip 5pt$\underline{\hbox{\tt#2}}$#3$\underline{\hbox{\tt#4}}$\hfill}\vskip-4pt
}
\def\laff #1 #2 #3 #4 {\hbox{{\footnotesize
$\underline{\hbox{\tt#1}}${\tt#2}$\underline{\hbox{\tt#3}}${}}}#4\hskip 4pt
}
\def\haff #1 #2 #3 #4 {\hbox{\footnotesize\tt\hbox to 13pt{\hfill#1}
\hskip 5pt$\underline{\hbox{\tt#2}}$#3$\underline{\hbox{\tt#4}}$}
}
\def\hraff #1 #2 #3 #4 {\hbox{\footnotesize\tt\hbox to 13pt{\hfill{\Rr{#1}}}
\hskip 5pt$\underline{\hbox{\tt#2}}$#3$\underline{\hbox{\tt#4}}$}
}
\newdimen\tabruleli\tabruleli=320pt
\newdimen\tabrulecol\tabrulecol=65pt

\subsection{The rules for the tracks}
\label{srtrack}

    Figure~\ref{elemtrack} shows us a single element of the track. Figure~\ref{ftracks}
shows us how to assemble elements as illustrated in Figure~\ref{elemtrack} in order
to constitute tracks. In Figure~\ref{ftracks}, we can see two tracks a part of which
goes around the central cell. We consider the following tracks:
to the right-hand side of~0(0), in yellow: the cells 17(8), 4(8), 3(8), 2(8), 1(7), 
1(6), 1(5), 1(4), 2(4), 5(3), 4(3) and 13(3); to the left-hand side of~0(0), in green: 
32(3), 9(3), 2(3), 1(2), 1(1), 2(1), 6(1) and 21(1). With this order, both tracks can be 
seen as a traversal in a clockwise motion. We shall look at both directions as suggested
by Table~\ref{rvoies} whose rules concern both tracks when a simple locomotive runs
over them.

   A close look at the tracks shows us at least two kinds of cells. There are 
three-milestoned cells as, for instance, 2(8) and 1(6), and four-milestoned ones
as, for instance, 4(8) and 1(4). Table~\ref{evsh} shows us that for the cells 2(8) 
and 1(6), the same rules were applied:
\vskip 5pt
\ligne{\hfill
\haff { 16} {W} {WBWWWBWB} {W} \hskip 10pt 
\hraff { 18} {W} {WBWWWBBB} {B} \hskip 10pt
\hraff { 24} {B} {WBWWWBWB} {W} \hskip 10pt
\haff { 30} {W} {BBWWWBWB} {W}  
\hfill}
\vskip 5pt
Indeed, in those cells, taking into account the numbering of the sides defined
in Section~\ref{scenar}, the milestones are neighbours 2, 6 and~8. Rule~16 is the 
conservative rule of the cell. Rule~\Rr{18} can see the locomotive through neighbour~7,
rule~\Rr{24} can see the locomotive in the cell so that its neighbourhood is the same as
that of rule~16, at last, rule~30 can see the locomotive in neighbour~1, witnessing that
the locomotive is now in the next cell of the track. Here and in the tables, 
a red number for a rule means that its new state is opposite to its current one.
\vskip 10pt
\vtop{
\ligne{\hfill
\includegraphics[scale=1]{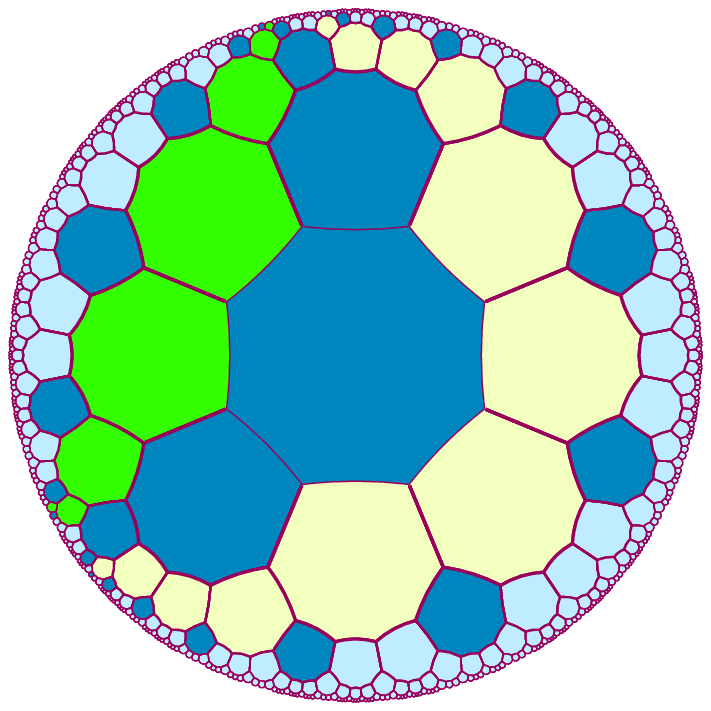}
\hfill}
\vspace{-15pt}
\ligne{\hfill
\vtop{\leftskip 0pt\parindent 0pt\hsize=230pt
\begin{fig}\label{ftracks}
\leurre
Assembling elements of the tracks in order to construct tracks, idle configuration.
\end{fig}
}
\hfill}
}

Table~\ref{evsh} shows us that for the cells~4(8) and~1(4), the rules are:
\vskip 5pt
\ligne{\hfill
\haff {  4} {W} {WBWWBWBB} {W} \hskip 10pt
\hraff { 27} {W} {WBWWBBBB} {B} \hskip 10pt
\hraff { 19} {B} {WBWWBWBB} {W} \hskip 10pt
\haff { 25} {W} {BBWWBWBB} {W} 
\hfill}

\ligne{\hfill 
\vtop{\leftskip 0pt\parindent 0pt\hsize=\tabruleli  
\begin{tab}\label{rvoies}
\leurre
Rules managing the motion of a simple locomotive on the tracks.
\end{tab}
\vskip-2pt
\trep
\vskip 8pt
\ligne{\hfill clockwise motion\hfill}
\ligne{\hfill   
\vtop{\leftskip 0pt\parindent 0pt\hsize=\tabrulecol  
\aff {  1} {W} {WWWWWWWW} {W} 
\aff {  2} {B} {WWWWWWWW} {B} 
\aff {  3} {B} {WWBWWBWW} {B} 
\aff {  4} {W} {WBWWBWBB} {W} 
\aff {  5} {W} {WBWWWWWW} {W} 
\aff {  6} {W} {WWWWWWWB} {W} 
\aff {  7} {W} {BWWWWWWW} {W} 
\aff {  8} {W} {WBWWWWWB} {W} 
\aff {  9} {B} {BWWBWWWW} {B} 
\aff { 10} {B} {BWWWWBWW} {B} 
}\hskip 10pt
\vtop{\leftskip 0pt\parindent 0pt\hsize=\tabrulecol  
\aff { 11} {W} {WBBWBWWB} {W}
\aff { 12} {W} {WBWBWWWB} {W}
\aff { 13} {W} {BWWWWWWB} {W}
\aff { 14} {B} {BWWWWWWW} {B}
\aff { 15} {W} {WWWWBWBW} {W}
\aff { 16} {W} {WBWWWBWB} {W}
\aff { 17} {B} {BWWWBBWW} {B}
\raff { 18} {W} {WBWWWBBB} {B}
\raff { 19} {B} {WBWWBWBB} {W}
\aff { 20} {B} {BBWWWBWW} {B}
}\hskip 10pt
\vtop{\leftskip 0pt\parindent 0pt\hsize=\tabrulecol  
\aff { 21} {W} {BBWWWWWW} {W}
\aff { 22} {W} {BBWBWWWB} {W}
\aff { 23} {B} {BWWBWBWW} {B}
\raff { 24} {B} {WBWWWBWB} {W}
\aff { 25} {W} {BBWWBWBB} {W}
\aff { 26} {B} {WBWWWWWW} {B}
\raff { 27} {W} {WBWWBBBB} {B}
\aff { 28} {B} {WWWWWWWB} {B}
\aff { 29} {B} {BWBWWBWW} {B}
\aff { 30} {W} {BBWWWBWB} {W}
}\hskip 10pt
\vtop{\leftskip 0pt\parindent 0pt\hsize=\tabrulecol  
\aff { 31} {B} {WBBWWBWW} {B} 
\aff { 32} {B} {WWBWWBWB} {B} 
\aff { 33} {B} {WWWWWWBW} {B} 
\aff { 34} {B} {WWBWWBBW} {B} 
\aff { 35} {B} {BWWBWWWB} {B} 
\aff { 36} {B} {BWWBWWBW} {B} 
\aff { 37} {B} {BWWBBWWW} {B} 
\aff { 38} {B} {BWWWWBWB} {B} 
}
\hfill}
\vskip 8pt
\ligne{\hfill counter-clockwise motion\hfill}
\ligne{\hfill   
\vtop{\leftskip 0pt\parindent 0pt\hsize=\tabrulecol  
\raff { 39} {B} {WBBWBWWB} {W} 
\raff { 40} {W} {WBBBWWWB} {B} 
}\hskip 10pt
\vtop{\leftskip 0pt\parindent 0pt\hsize=\tabrulecol  
\aff { 41} {W} {BBBWBWWB} {W} 
}\hskip 10pt
\vtop{\leftskip 0pt\parindent 0pt\hsize=\tabrulecol  
\raff { 42} {B} {WBWBWWWB} {W} 
}\hskip 10pt
\vtop{\leftskip 0pt\parindent 0pt\hsize=\tabrulecol  
\raff { 43} {W} {WBBBBWWB} {B} 
}
\hfill}  
\vskip 9pt
\trfn
\vskip 8pt
}
\hfill}

The milestones of the cell are in neighbours~2, 5, 7 and~8. The conservative rule
of the cell is rule~4. Rule~\Rr{27} can see the arriving locomotive through neighbour~6,
rule~\Rr{19} can see it in the cell, so that its neighbouring is also that of rule~4.
At last, rule~25 can see the locomotive in neighbour~1, witnessing that it is now
in the next cell of the track. Table~\ref{evshb} shows us that the same rules
for four-milestoned cells are used as in the motion on the left-hand side,
still in the clockwise motion.


\newdimen\largez\largez=20pt
\def\HH#1{\hbox to\largez{\hfill #1\hfill}}
\ligne{\hfill
\vtop{\hsize=240pt
\begin{tab}\label{evsh}\leurre
Execution of the rules $1$ up to~$38$: motion along the tracks in the clockwise
direction.
\end{tab}
\vskip 2pt
\trep
\vskip 8pt
\ligne{\hfill\HH{}
\HH{{4$_8$} }\HH{{3$_8$} }\HH{{2$_8$} }\HH{{1$_7$} }\HH{{1$_6$} }
\HH{{1$_5$} }\HH{{1$_4$} }\HH{{2$_4$} }\HH{{5$_3$} }\HH{{4$_3$} }\hfill}
\ligne{\hfill\HH{1}
\HH{25}\HH{\Rr{24}}\HH{\Rr{18}}\HH{4}\HH{16}
\HH{16}\HH{4}\HH{16}\HH{16}\HH{4}\hfill}
\ligne{\hfill\HH{2}
\HH{4}\HH{30}\HH{\Rr{24}}\HH{\Rr{27}}\HH{16}
\HH{16}\HH{4}\HH{16}\HH{16}\HH{4}\hfill}
\ligne{\hfill\HH{3}
\HH{4}\HH{16}\HH{30}\HH{\Rr{19}}\HH{\Rr{18}}
\HH{16}\HH{4}\HH{16}\HH{16}\HH{4}\hfill}
\ligne{\hfill\HH{4}
\HH{4}\HH{16}\HH{16}\HH{25}\HH{\Rr{24}}
\HH{\Rr{18}}\HH{4}\HH{16}\HH{16}\HH{4}\hfill}
\ligne{\hfill\HH{5}
\HH{4}\HH{16}\HH{16}\HH{4}\HH{30}
\HH{\Rr{24}}\HH{\Rr{27}}\HH{16}\HH{16}\HH{4}\hfill}
\ligne{\hfill\HH{6}
\HH{4}\HH{16}\HH{16}\HH{4}\HH{16}
\HH{30}\HH{\Rr{19}}\HH{\Rr{18}}\HH{16}\HH{4}\hfill}
\ligne{\hfill\HH{7}
\HH{4}\HH{16}\HH{16}\HH{4}\HH{16}
\HH{16}\HH{25}\HH{\Rr{24}}\HH{\Rr{18}}\HH{4}\hfill}
\ligne{\hfill\HH{8}
\HH{4}\HH{16}\HH{16}\HH{4}\HH{16}
\HH{16}\HH{4}\HH{30}\HH{\Rr{24}}\HH{\Rr{27}}\hfill}
\ligne{\hfill\HH{9}
\HH{4}\HH{16}\HH{16}\HH{4}\HH{16}
\HH{16}\HH{4}\HH{16}\HH{30}\HH{\Rr{19}}\hfill}
\vskip 9pt
\trfn
\vskip 15pt
}
\hfill}

\ligne{
\vtop{\hsize=160pt
\begin{tab}\label{evshb}\leurre
Execution of the rules $1$ up to~$38$: in the clockwise direction, on the left-hand side
of~$0(0)$.
\end{tab}
\vskip 2pt
\trep
\vskip 8pt
\ligne{\hfill\HH{}
\HH{{9$_3$} }\HH{{2$_3$} }\HH{{1$_2$} }\HH{{1$_1$} }\HH{{2$_1$} }\HH{{6$_1$} }
\hfill}
\ligne{\hfill\HH{1}
\HH{22}\HH{\Rr{19}}\HH{\Rr{27}}\HH{4}\HH{4}\HH{4}\hfill}
\ligne{\hfill\HH{2}
\HH{12}\HH{25}\HH{\Rr{19}}\HH{\Rr{27}}\HH{4}\HH{4}\hfill}
\ligne{\hfill\HH{3}
\HH{12}\HH{4}\HH{25}\HH{\Rr{19}}\HH{\Rr{27}}\HH{4}\hfill}
\ligne{\hfill\HH{4}
\HH{12}\HH{4}\HH{4}\HH{25}\HH{\Rr{19}}\HH{\Rr{27}}\hfill}
\vskip 9pt
\trfn
\vskip 15pt
}
\hskip 20pt
\vtop{\hsize=160pt
\begin{tab}\label{evsahb}\leurre
Execution of the rules $1$ up to~$38$: in the counter-clockwise direction, on the 
right-hand side of~$0(0)$.
\end{tab}
\vskip 2pt
\trep
\vskip 8pt
\ligne{\hfill\HH{}
\HH{{6$_1$} }\HH{{2$_1$} }\HH{{1$_1$} }\HH{{1$_2$} }\HH{{2$_3$} }\HH{{9$_3$} }
\hfill}
\ligne{\hfill\HH{1}
\HH{41}\HH{\Rr{39}}\HH{\Rr{43}}\HH{11}\HH{11}\HH{16}\hfill}
\ligne{\hfill\HH{2}
\HH{11}\HH{41}\HH{\Rr{39}}\HH{\Rr{43}}\HH{11}\HH{16}\hfill}
\ligne{\hfill\HH{3}
\HH{11}\HH{11}\HH{41}\HH{\Rr{39}}\HH{\Rr{43}}\HH{16}\hfill}
\ligne{\hfill\HH{4}
\HH{11}\HH{11}\HH{11}\HH{41}\HH{\Rr{39}}\HH{\Rr{18}}\hfill}
\vskip 9pt
\trfn
\vskip 15pt
}
\hfill}

Now, Table~\ref{evsah} and~\ref{evsahb} show us that other rules occur when the
motion happens on the same tracks but in the opposite direction: counter-clockwise.
For three-milestoned cells we have now the following rules:
\vskip 5pt
\ligne{\hfill
\haff { 12} {W} {WBWBWWWB} {W} \hskip 10pt
\hraff { 40} {W} {WBBBWWWB} {B} \hskip 10pt
\hraff { 42} {B} {WBWBWWWB} {W} \hskip 10pt
\haff { 22} {W} {BBWBWWWB} {W} 
\hfill}
\vskip 5pt
Note that the milestones are now in neighbours~2, 4 and~8 instead of ~2, 6 and~8: the 
change of side~1 explains the new situation. The rules show that the locomotive is 
successively seen in neighbours~3, 0 and~1 as expected, where neighbour~0 is the cell 
itself.

For four-milestoned cells the rules are now:
\vskip 5pt
\ligne{\hfill
\haff { 11} {W} {WBBWBWWB} {W} \hskip 10pt
\hraff { 43} {W} {WBBBBWWB} {B} \hskip 10pt
\hraff { 39} {B} {WBBWBWWB} {W} \hskip 10pt
\haff { 41} {W} {BBBWBWWB} {W} 
\hfill}
\vskip 5pt
The milestones are now in neighbours~2, 3, 5 and~8 instead of 2, 5, 7 and~8 for the same
reason. The rules indicate that the locomotive is successively seen in neighbours~4,
0 and~1.

\ligne{\hfill
\vtop{\hsize=240pt
\begin{tab}\label{evsah}
Execution of the rules $1$ up to~$43$: the locomotive on the tracks in the 
counter-clockwise direction.
\end{tab}
\vskip 2pt
\trep
\vskip 8pt
\ligne{\hfill\HH{}
\HH{{4$_3$} }\HH{{5$_3$} }\HH{{2$_4$} }\HH{{1$_4$} }\HH{{1$_5$} }
\HH{{1$_6$} }\HH{{1$_7$} }\HH{{2$_8$} }\HH{{3$_8$} }\HH{{4$_8$} }\hfill}
\ligne{\hfill\HH{1}
\HH{41}\HH{\Rr{42}}\HH{\Rr{40}}\HH{11}\HH{12}
\HH{12}\HH{11}\HH{12}\HH{12}\HH{11}\hfill}
\ligne{\hfill\HH{2}
\HH{11}\HH{22}\HH{\Rr{42}}\HH{\Rr{43}}\HH{12}
\HH{12}\HH{11}\HH{12}\HH{12}\HH{11}\hfill}
\ligne{\hfill\HH{3}
\HH{11}\HH{12}\HH{22}\HH{\Rr{39}}\HH{\Rr{40}}
\HH{12}\HH{11}\HH{12}\HH{12}\HH{11}\hfill}
\ligne{\hfill\HH{4}
\HH{11}\HH{12}\HH{12}\HH{41}\HH{\Rr{42}}
\HH{\Rr{40}}\HH{11}\HH{12}\HH{12}\HH{11}\hfill}
\ligne{\hfill\HH{5}
\HH{11}\HH{12}\HH{12}\HH{11}\HH{22}
\HH{\Rr{42}}\HH{\Rr{43}}\HH{12}\HH{12}\HH{11}\hfill}
\ligne{\hfill\HH{6}
\HH{11}\HH{12}\HH{12}\HH{11}\HH{12}
\HH{22}\HH{\Rr{39}}\HH{\Rr{40}}\HH{12}\HH{11}\hfill}
\ligne{\hfill\HH{7}
\HH{11}\HH{12}\HH{12}\HH{11}\HH{12}
\HH{12}\HH{41}\HH{\Rr{42}}\HH{\Rr{40}}\HH{11}\hfill}
\ligne{\hfill\HH{8}
\HH{11}\HH{12}\HH{12}\HH{11}\HH{12}
\HH{12}\HH{11}\HH{22}\HH{\Rr{42}}\HH{\Rr{43}}\hfill}
\ligne{\hfill\HH{9}
\HH{11}\HH{12}\HH{12}\HH{11}\HH{12}
\HH{12}\HH{11}\HH{12}\HH{22}\HH{\Rr{39}}\hfill}
\vskip 9pt
\trfn
\vskip 15pt
}
\hfill}

\vtop{
\ligne{\hfill
\includegraphics[scale=0.55]{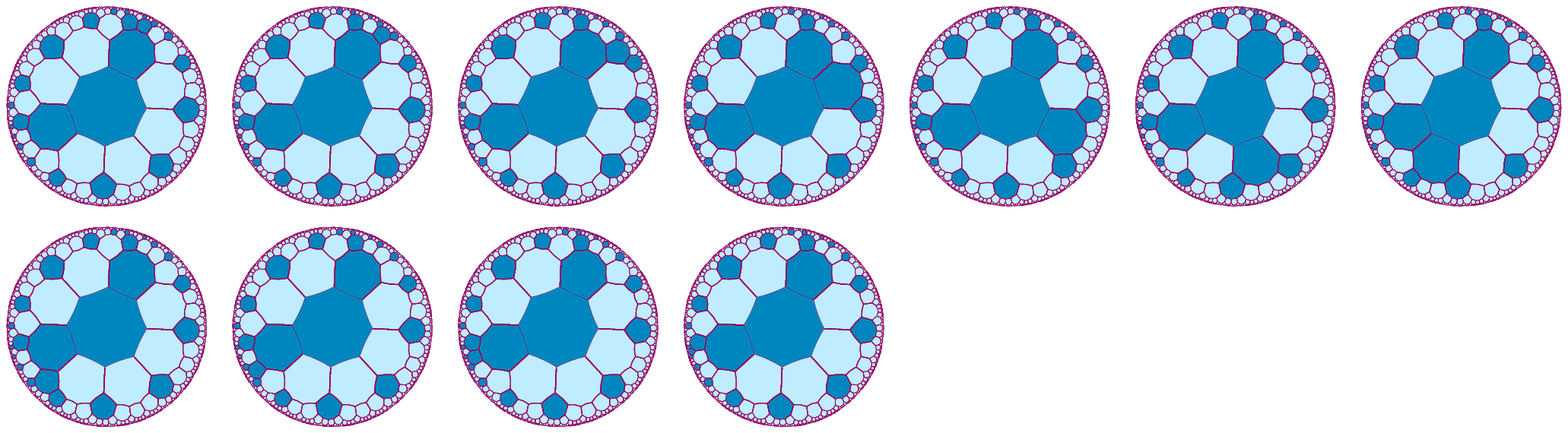}
\hfill}
\vspace{-20pt}
\ligne{\hfill
\includegraphics[scale=0.55]{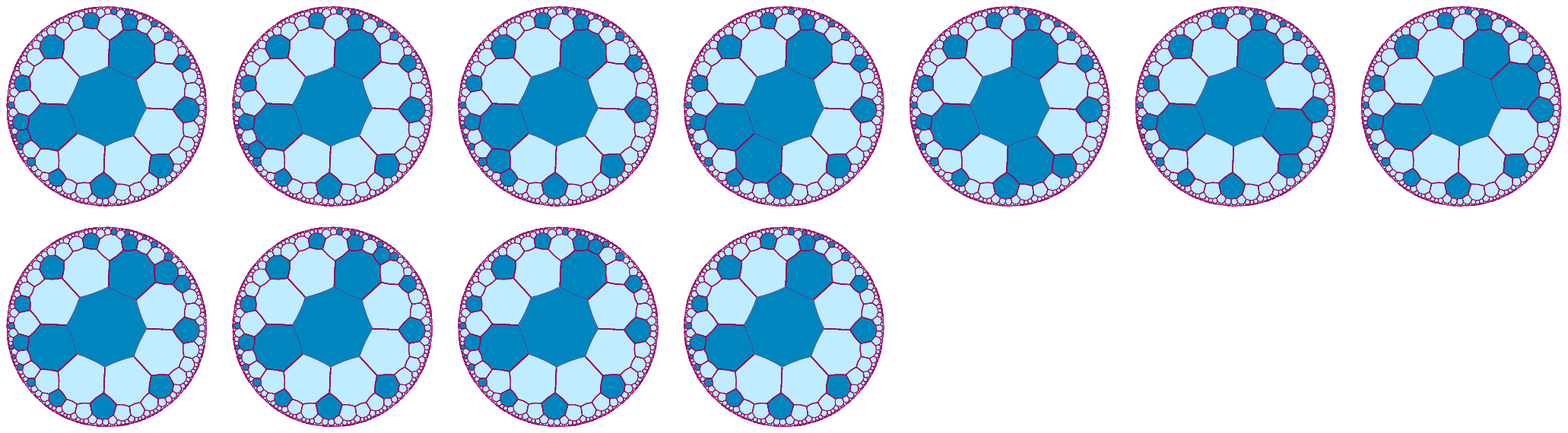}
\hfill}
\vspace{-15pt}
\ligne{\hfill
\vtop{\leftskip 0pt\parindent 0pt\hsize=300pt
\begin{fig}\label{fvs}
\leurre
Illustration of the motion of a simple locomotive in a clockwise motion, above,
and in a counterclockwise one, below.
\end{fig}
}
\hfill}
}

Figure~\ref{fvs} illustrates the executions given by Tables~\ref{evsh} and~\ref{evsah}.
The motion of the simple locomotive is performed on the tracks which
lie on the right-hand side of the central cell in Figure~\ref{ftracks}. In the first
two rows the motion is clockwise. In the last two rows, the motion is counterclockwise
on the same set of cells.
Figure~\ref{fvsb} illustrates the executions given by Tables~\ref{evshb} 
and~\ref{evsahb} where the simple locomotive runs over the tracks illustrated by the 
left-hand side of cell~0(0) in Figure~\ref{ftracks}. In the first row, the motion is 
clockwise, on the second one, it is counterclockwise.

\vtop{
\ligne{\hskip 25pt
\includegraphics[scale=0.55]{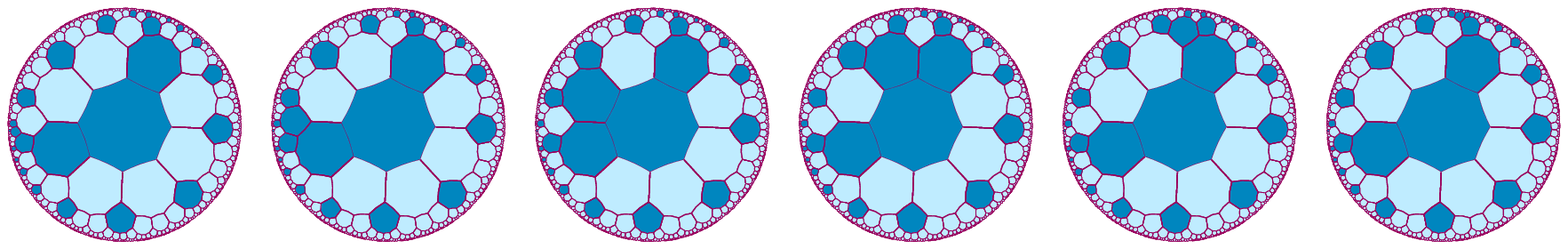}
\hfill}
\vspace{-20pt}
\ligne{\hskip 25pt
\includegraphics[scale=0.55]{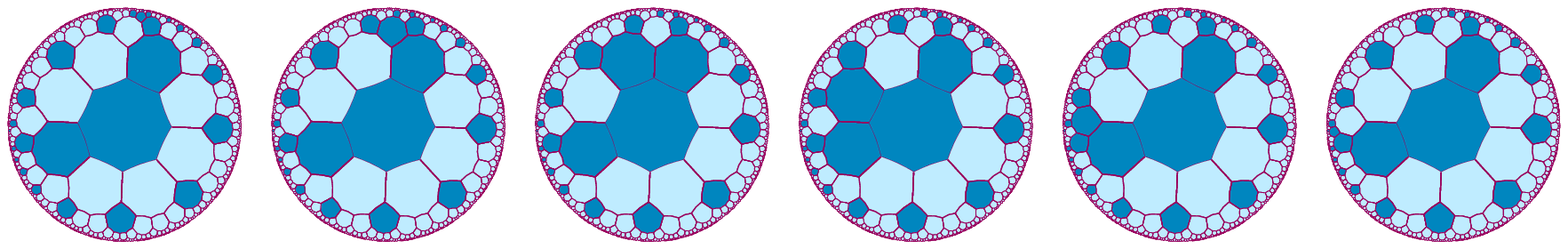}
\hfill}
\vspace{-15pt}
\ligne{\hfill
\vtop{\leftskip 0pt\parindent 0pt\hsize=300pt
\begin{fig}\label{fvsb}
\leurre
Illustration of the motion of a simple locomotive on tracks in an counterclockwise motion.
\end{fig}
}
\hfill}
}
\vskip 10pt
Let us now look at the motion of a double locomotive on the same tracks, again in both
directions. Table~\ref{rvoied} gives the rules concerning that situation.

\ligne{\hfill 
\vtop{\leftskip 0pt\parindent 0pt\hsize=\tabruleli  
\begin{tab}\label{rvoied}
\leurre
Rules for the motion of a double locomotive on the tracks.
\end{tab}
\vskip-2pt
\trep
\vskip 8pt
\ligne{\hfill   
\vtop{\leftskip 0pt\parindent 0pt\hsize=\tabrulecol  
\aff { 44} {B} {BWWBBBWW} {B}
\aff { 45} {B} {WBWWWBBB} {B}
\raff { 46} {B} {BBWWBWBB} {W}
\aff { 47} {B} {BBWWWWWW} {B}
\aff { 48} {B} {BWBBWBWW} {B}
\raff { 49} {B} {BBWWWBWB} {W}
\aff { 50} {B} {WBWWBBBB} {B}
}\hskip 10pt
\vtop{\leftskip 0pt\parindent 0pt\hsize=\tabrulecol  
\aff { 51} {B} {BWWWWWWB} {B} 
\aff { 52} {B} {BBBWWBWW} {B} 
\aff { 53} {B} {BWBWWBWB} {B} 
\aff { 54} {B} {WWBWWBBB} {B} 
\aff { 55} {B} {WWWWWWBB} {B} 
\aff { 56} {B} {BWWBWWBB} {B} 
\aff { 57} {B} {BWWBWBBW} {B} 
}\hskip 10pt
\vtop{\leftskip 0pt\parindent 0pt\hsize=\tabrulecol  
\raff { 58} {B} {BBBWBWWB} {W} 
\aff { 59} {B} {WBBBWWWB} {B} 
\raff { 60} {B} {BBWBWWWB} {W} 
\aff { 61} {B} {WBBBBWWB} {B} 
\aff { 62} {W} {BBWWWWWB} {W} 
\aff { 63} {B} {BWBBWWWW} {B} 
\aff { 64} {B} {WWBWBBWW} {B} 
}\hskip 10pt
\vtop{\leftskip 0pt\parindent 0pt\hsize=\tabrulecol  
\aff { 65} {B} {BBWBWWWW} {B} 
\aff { 66} {B} {WWBBWBWW} {B} 
\aff { 67} {B} {BWWWWBBW} {B} 
\aff { 68} {B} {BBBBWWWW} {B} 
\aff { 69} {B} {WWBBBBWW} {B} 
\aff { 70} {B} {BWWWWBBB} {B} 
}
\hfill} 
\vskip 9pt
\trfn
\vskip 8pt
} 
\hfill}

As can be seen on Tables~\ref{evdh}, \ref{evdhb}, \ref{evdah} and~\ref{evdahb},
the rules for a double locomotive on the tracks makes use of the already mentioned
rules. To these ones, we have to append rules which are induced by the structure
of the double locomotive. This has an impact on the motion rules and also on several
milestones. From Table~\ref{evdh}, we can see that for the three-milestoned cells,
the sequence of rules is not 16, \Rr{18}, \Rr{24} and~30. Rule~\Rr{24} is no more used 
and it is replaced by the occurrence of two rules:
\vskip 5pt
\ligne{\hfill
\haff { 45} {B} {WBWWWBBB} {B} \hskip 20pt
\hraff { 49} {B} {BBWWWBWB} {W} 
\hfill}
\vskip 5pt
Indeed, rule~\Rr{18} applies as the cell can only see the front of the arriving locomotive.
When the front of the locomotive is in the cell, its rear is in neighbour~7 so that
rule~\Rr{24} cannot be applied. But rule~45 does apply: it can see the front of the
locomotive in the cell itself and the rear in neighbour~7. Next, rule~\Rr{49} applies:
the front of the locomotive has been absorbed by the next cell: it is in 
the neighbour~1 of the cell under examination. The rear is still here, so it must
leave the cell, which is performed by rule~\Rr{49}. 
Now, when the rear of the 
locomotive is in neighbour~1, the cell cannot see the front of the locomotive. 
Accordingly, rule~30 again applies, witnessing that the locomotive left the cell.

\ligne{\hfill
\vtop{\hsize=240pt
\begin{tab}\label{evdh}
Execution of the rules $1$ up to~$70$: motion in the clockwise direction
for a double locomotive.
\end{tab}
\vskip 2pt
\trep
\vskip 8pt
\ligne{\hfill\HH{}
\HH{{4$_8$} }\HH{{3$_8$} }\HH{{2$_8$} }\HH{{1$_7$} }\HH{{1$_6$} }
\HH{{1$_5$} }\HH{{1$_4$} }\HH{{2$_4$} }\HH{{5$_3$} }\HH{{4$_3$} }\hfill}
\ligne{\hfill\HH{1}
\HH{25}\HH{\Rr{49}}\HH{45}\HH{\Rr{27}}\HH{16}
\HH{16}\HH{4}\HH{16}\HH{16}\HH{4}\hfill}
\ligne{\hfill\HH{2}
\HH{4}\HH{30}\HH{\Rr{49}}\HH{50}\HH{\Rr{18}}
\HH{16}\HH{4}\HH{16}\HH{16}\HH{4}\hfill}
\ligne{\hfill\HH{3}
\HH{4}\HH{16}\HH{30}\HH{\Rr{46}}\HH{45}
\HH{\Rr{18}}\HH{4}\HH{16}\HH{16}\HH{4}\hfill}
\ligne{\hfill\HH{4}
\HH{4}\HH{16}\HH{16}\HH{25}\HH{\Rr{49}}
\HH{45}\HH{\Rr{27}}\HH{16}\HH{16}\HH{4}\hfill}
\ligne{\hfill\HH{5}
\HH{4}\HH{16}\HH{16}\HH{4}\HH{30}
\HH{\Rr{49}}\HH{50}\HH{\Rr{18}}\HH{16}\HH{4}\hfill}
\ligne{\hfill\HH{6}
\HH{4}\HH{16}\HH{16}\HH{4}\HH{16}
\HH{30}\HH{\Rr{46}}\HH{45}\HH{\Rr{18}}\HH{4}\hfill}
\ligne{\hfill\HH{7}
\HH{4}\HH{16}\HH{16}\HH{4}\HH{16}
\HH{16}\HH{25}\HH{\Rr{49}}\HH{45}\HH{\Rr{27}}\hfill}
\ligne{\hfill\HH{8}
\HH{4}\HH{16}\HH{16}\HH{4}\HH{16}
\HH{16}\HH{4}\HH{30}\HH{\Rr{49}}\HH{50}\hfill}
\vskip 9pt
\trfn
\vskip 15pt
}
\hfill}

\ligne{\hfill
\vtop{\hsize=160pt
\begin{tab}\label{evdhb}
Execution of the rules $1$ up to~$70$: clockwise, double locomotive,
on the left-hand side of $0(0)$.
\end{tab}
\vskip 2pt
\trep
\vskip 8pt
\ligne{\hfill\HH{}
\HH{{9$_3$} }\HH{{2$_3$} }\HH{{1$_2$} }\HH{{1$_1$} }\HH{{2$_1$} }\HH{{6$_1$} }
\hfill}
\ligne{\hfill\HH{1}
\HH{22}\HH{\Rr{46}}\HH{50}\HH{\Rr{27}}\HH{4}\HH{4}\hfill}
\ligne{\hfill\HH{2}
\HH{12}\HH{25}\HH{\Rr{46}}\HH{50}\HH{\Rr{27}}\HH{4}\hfill}
\ligne{\hfill\HH{3}
\HH{12}\HH{4}\HH{25}\HH{\Rr{46}}\HH{50}\HH{\Rr{27}}\hfill}
\vskip 9pt
\trfn
\vskip 15pt
}
\hskip 20pt
\vtop{\hsize=160pt
\begin{tab}\label{evdahb}
Execution of the rules $1$ up to~$70$: counterclockwise, double locomotive,
on the left-hand side of $0(0)$.
\end{tab}
\vskip 2pt
\trep
\vskip 8pt
\ligne{\hfill\HH{}
\HH{{6$_1$} }\HH{{2$_1$} }\HH{{1$_1$} }\HH{{1$_2$} }\HH{{2$_3$} }\HH{{9$_3$} }
\hfill}
\ligne{\hfill\HH{1}
\HH{41}\HH{\Rr{58}}\HH{61}\HH{\Rr{43}}\HH{11}\HH{16}\hfill}
\ligne{\hfill\HH{2}
\HH{11}\HH{41}\HH{\Rr{58}}\HH{61}\HH{\Rr{43}}\HH{16}\hfill}
\ligne{\hfill\HH{3}
\HH{11}\HH{11}\HH{41}\HH{\Rr{58}}\HH{61}\HH{\Rr{18}}\hfill}
\vskip 9pt
\trfn
\vskip 15pt
}
\hfill}

\ligne{\hfill
\vtop{\hsize=240pt
\begin{tab}\label{evdah}\leurre
Execution of the rules~$1$ up to~$70$ for a double locomotive in a counterclockwise
motion.
\end{tab}
\vskip-2pt
\trep
\vskip 8pt
\ligne{\hfill\HH{}
\HH{{4$_3$} }\HH{{5$_3$} }\HH{{2$_4$} }\HH{{1$_4$} }\HH{{1$_5$} }
\HH{{1$_6$} }\HH{{1$_7$} }\HH{{2$_8$} }\HH{{3$_8$} }\HH{{4$_8$} }\hfill}
\ligne{\hfill\HH{1}
\HH{41}\HH{\Rr{60}}\HH{59}\HH{\Rr{43}}\HH{12}
\HH{12}\HH{11}\HH{12}\HH{12}\HH{11}\hfill}
\ligne{\hfill\HH{2}
\HH{11}\HH{22}\HH{\Rr{60}}\HH{61}\HH{\Rr{40}}
\HH{12}\HH{11}\HH{12}\HH{12}\HH{11}\hfill}
\ligne{\hfill\HH{3}
\HH{11}\HH{12}\HH{22}\HH{\Rr{58}}\HH{59}
\HH{\Rr{40}}\HH{11}\HH{12}\HH{12}\HH{11}\hfill}
\ligne{\hfill\HH{4}
\HH{11}\HH{12}\HH{12}\HH{41}\HH{\Rr{60}}
\HH{59}\HH{\Rr{43}}\HH{12}\HH{12}\HH{11}\hfill}
\ligne{\hfill\HH{5}
\HH{11}\HH{12}\HH{12}\HH{11}\HH{22}
\HH{\Rr{60}}\HH{61}\HH{\Rr{40}}\HH{12}\HH{11}\hfill}
\ligne{\hfill\HH{6}
\HH{11}\HH{12}\HH{12}\HH{11}\HH{12}
\HH{22}\HH{\Rr{58}}\HH{59}\HH{\Rr{40}}\HH{11}\hfill}
\ligne{\hfill\HH{7}
\HH{11}\HH{12}\HH{12}\HH{11}\HH{12}
\HH{12}\HH{41}\HH{\Rr{60}}\HH{59}\HH{\Rr{43}}\hfill}
\ligne{\hfill\HH{8}
\HH{11}\HH{12}\HH{12}\HH{11}\HH{12}
\HH{12}\HH{11}\HH{22}\HH{\Rr{60}}\HH{61}\hfill}
\vskip 9pt
\trfn
\vskip 8pt
}
\hfill}

We can perform a similar analysis for the cells with four milestones:
\vskip 5pt
\ligne{\hfill
\haff { 50} {B} {WBWWBBBB} {B} \hskip 20pt
\hraff { 46} {B} {BBWWBWBB} {W} 
\hfill}
\vskip 5pt
In that case, where rules~4, \Rr{27}, \Rr{19}, and~25 apply for a simple locomotive, here
rule~\Rr{19} is no more used and is replaced by rules~50 and~\Rr{46} which then allows 
rule~25 to be again applied.

\vtop{
\ligne{\hfill
\includegraphics[scale=0.55]{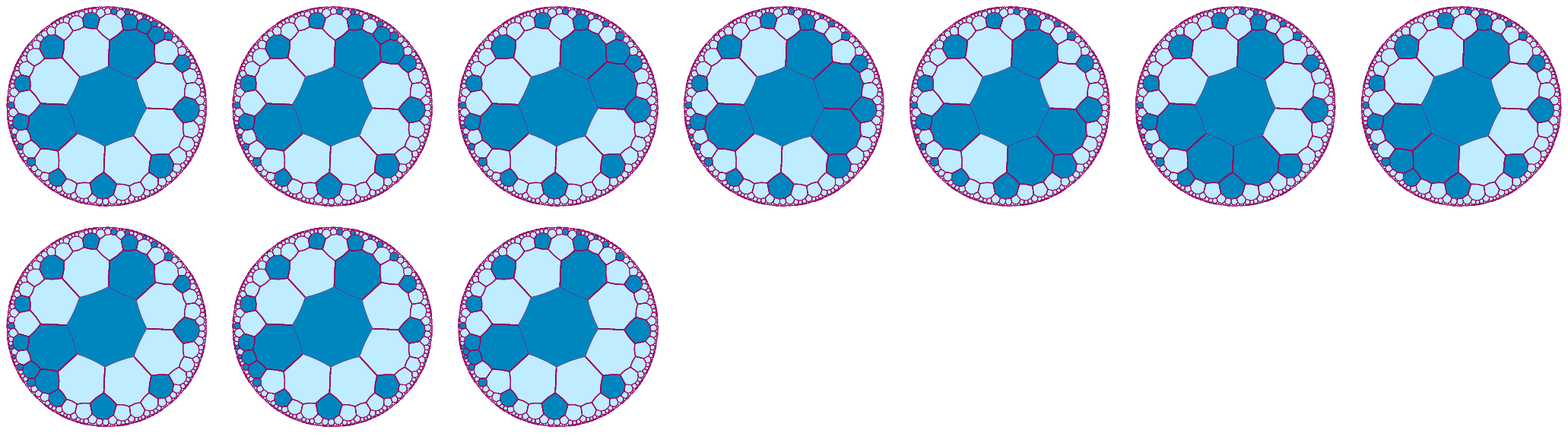}
\hfill}
\vspace{-20pt}
\ligne{\hfill
\includegraphics[scale=0.55]{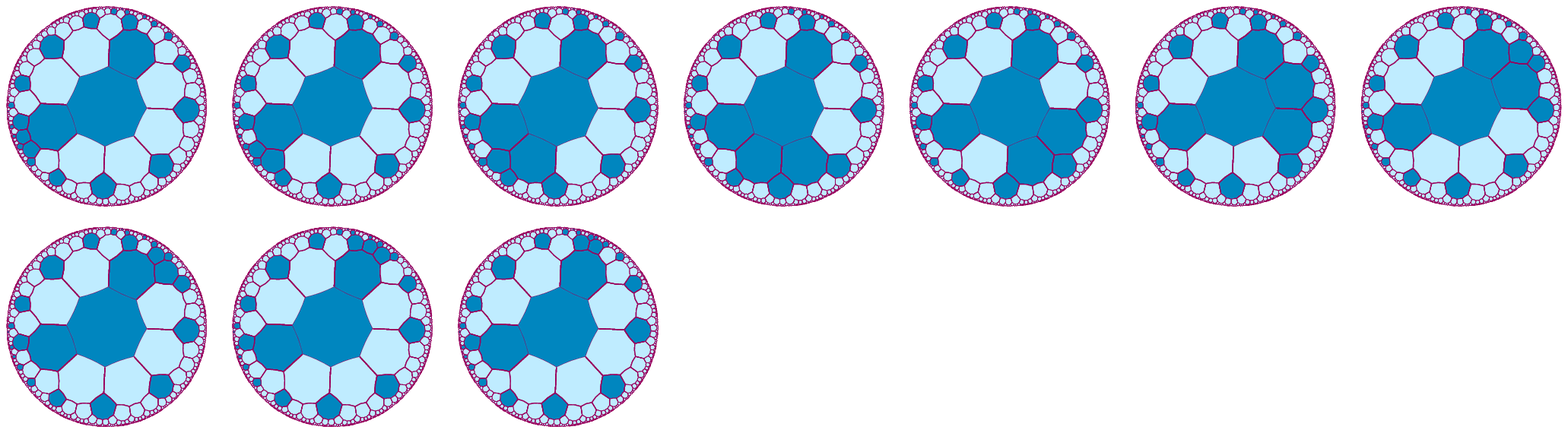}
\hfill}
\vspace{-15pt}
\ligne{\hfill
\vtop{\leftskip 0pt\parindent 0pt\hsize=300pt
\begin{fig}\label{fvd}
\leurre
Illustration of the motion of a double locomotive in a clockwise motion, above,
and in a counterclockwise one, below.
\end{fig}
}
\hfill}
}

\vtop{
\ligne{\hskip 25pt
\includegraphics[scale=0.55]{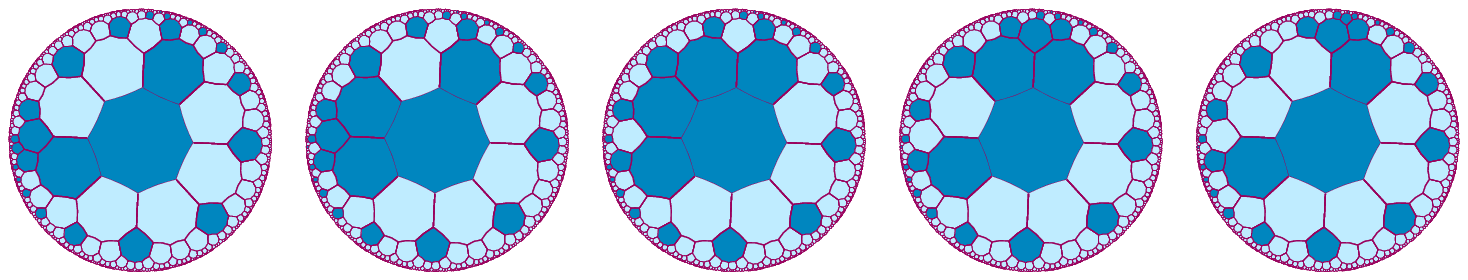}
\hfill}
\vspace{-20pt}
\ligne{\hskip 25pt
\includegraphics[scale=0.55]{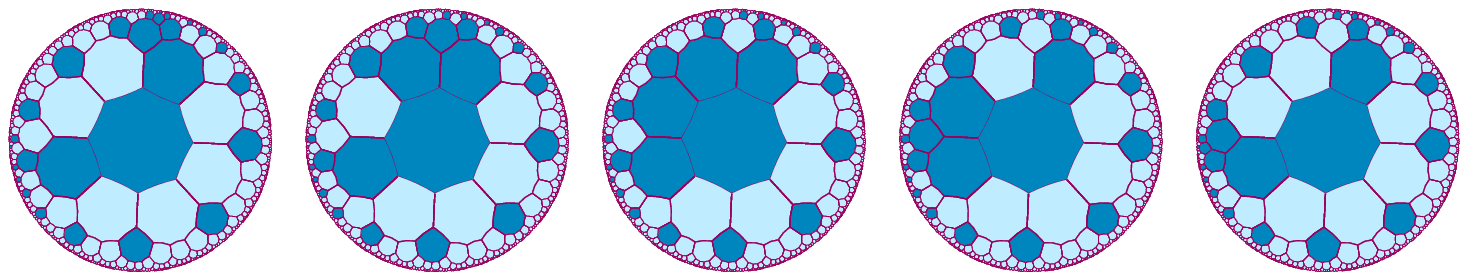}
\hfill}
\vspace{-15pt}
\ligne{\hfill
\vtop{\leftskip 0pt\parindent 0pt\hsize=300pt
\begin{fig}\label{fvdb}
\leurre
Illustration of the motion of a double locomotive in a clockwise motion, above,
and in a counterclockwise one, below.
\end{fig}
}
\hfill}
}

We noticed for a simple locomotive that the counterclockwise motion involved
new rules with respect to those managing a clockwise motion. The same occurs for the
double locomotive. In a three-milestoned cell, instead of the sequence of rules 12, 
\Rr{40}, \Rr{42} and~22 we have the rules 12, \Rr{40}, 59, \Rr{60}, and~22. Indeed, 
rule~\Rr{42} cannot be applied for the same reason that the locomotive now occupies two 
consecutive cells of the track. Instead rule~\Rr{42}, two new rules are involved, 
rule~59 and~\Rr{60}, see below and 
Table~\ref{rvoied}. Similarly, for a four-milestoned cell, the sequence of rules
11, \Rr{43}, \Rr{39} and~41 is replace by 11, \Rr{43}, 61, \Rr{58} and~41 where the 
two new rules~61 and~\Rr{58}, see
below and Table~\ref{rvoied}, replace rule~\Rr{39} which cannot be applied here.
\vskip 5pt
\ligne{\hfill
\haff { 59} {B} {WBBBWWWB} {B} \hskip 10pt
\hraff { 60} {B} {BBWBWWWB} {W} \hskip 30pt
\haff { 61} {B} {WBBBBWWB} {B} \hskip 10pt
\hraff { 58} {B} {BBBWBWWB} {W} 
\hfill}
\vskip 5pt
We conclude the sub-section with a remark on the rules managing the milestones. Several
of them witness the passage of the locomotive, either when it is simple or when it is
double. We shall consider three examples: the cells~1(8), 0(0) and 2(7), see 
Figure~\ref{ftracks}. Table~\ref{rmiles} gives the rules used for those cells
during the motion of the locomotive for the four cases: the simple locomotive
clockwise running, then for the same one counterclockwise running, then for the double
locomotive when it clockwise traverses the tracks and, at last, for the same locomotive 
when it counterclockwise traverses the tracks.

\ligne{\hfill  
\vtop{  
\begin{tab}\label{rmiles}
\leurre
Rules witnessing the passage of the locomotive. Symbols~$\curvearrowright$,
$\curvearrowleft$ indicate  clockwise motion, counterclockwise motions respectively.
\end{tab}
\vskip-2pt
\trep
\vskip 8pt
\ligne{\hskip 5pt 0(0): \hfill   
\vtop{\leftskip 0pt\parindent 0pt\hsize=320pt  
\ligne{\hfill  
\vtop{\leftskip 0pt\parindent 0pt\hsize=\tabrulecol  %
\ligne{\hfill simple, $\curvearrowright$\hfill}
}\hskip 10pt
\vtop{\leftskip 0pt\parindent 0pt\hsize=\tabrulecol  %
\ligne{\hfill simple, $\curvearrowleft$\hfill}
}\hskip 10pt
\vtop{\leftskip 0pt\parindent 0pt\hsize=\tabrulecol  %
\ligne{\hfill double, $\curvearrowright$\hfill}
}\hskip 10pt
\vtop{\leftskip 0pt\parindent 0pt\hsize=\tabrulecol  %
\ligne{\hfill double, $\curvearrowleft$\hfill}
}
\hfill}  
\ligne{\hfill 
\vtop{\leftskip 0pt\parindent 0pt\hsize=\tabrulecol  
\aff {  3} {B} {WWBWWBWW} {B}
\aff { 31} {B} {WBBWWBWW} {B}
\aff { 29} {B} {BWBWWBWW} {B}
\aff { 32} {B} {WWBWWBWB} {B}
\aff { 34} {B} {WWBWWBBW} {B}
\aff {  3} {B} {WWBWWBWW} {B}
}\hskip 10pt
\vtop{\leftskip 0pt\parindent 0pt\hsize=\tabrulecol  
\aff {  3} {B} {WWBWWBWW} {B}
\aff { 34} {B} {WWBWWBBW} {B}
\aff { 32} {B} {WWBWWBWB} {B}
\aff { 29} {B} {BWBWWBWW} {B}
\aff { 31} {B} {WBBWWBWW} {B}
\aff {  3} {B} {WWBWWBWW} {B}
}\hskip 10pt
\vtop{\leftskip 0pt\parindent 0pt\hsize=\tabrulecol  
\aff {  3} {B} {WWBWWBWW} {B}
\aff { 31} {B} {WBBWWBWW} {B}
\aff { 52} {B} {BBBWWBWW} {B}
\aff { 53} {B} {BWBWWBWB} {B}
\aff { 54} {B} {WWBWWBBB} {B}
\aff { 34} {B} {WWBWWBBW} {B}
\aff {  3} {B} {WWBWWBWW} {B}
}\hskip 10pt
\vtop{\leftskip 0pt\parindent 0pt\hsize=\tabrulecol  
\aff {  3} {B} {WWBWWBWW} {B}
\aff { 34} {B} {WWBWWBBW} {B}
\aff { 54} {B} {WWBWWBBB} {B}
\aff { 53} {B} {BWBWWBWB} {B}
\aff { 52} {B} {BBBWWBWW} {B}
\aff { 31} {B} {WBBWWBWW} {B}
\aff {  3} {B} {WWBWWBWW} {B}
}
\hfill} 
} 
\hfill} 
\vskip 9pt
\trfn
\vskip 5pt
\ligne{\hskip 5pt 1(8): \hfill   
\vtop{\leftskip 0pt\parindent 0pt\hsize=320pt  
\ligne{\hfill  
\vtop{\leftskip 0pt\parindent 0pt\hsize=\tabrulecol  %
\ligne{\hfill simple, $\curvearrowright$\hfill}
}\hskip 10pt
\vtop{\leftskip 0pt\parindent 0pt\hsize=\tabrulecol  %
\ligne{\hfill simple, $\curvearrowleft$\hfill}
}\hskip 10pt
\vtop{\leftskip 0pt\parindent 0pt\hsize=\tabrulecol  %
\ligne{\hfill double, $\curvearrowright$\hfill}
}\hskip 10pt
\vtop{\leftskip 0pt\parindent 0pt\hsize=\tabrulecol  %
\ligne{\hfill double, $\curvearrowleft$\hfill}
}
\hfill}  
\ligne{\hfill  
\vtop{\leftskip 0pt\parindent 0pt\hsize=\tabrulecol  %
\aff { 17} {B} {BWWWBBWW} {B}
\aff { 23} {B} {BWWBWBWW} {B}
\aff { 29} {B} {BWBWWBWW} {B}
\aff { 20} {B} {BBWWWBWW} {B}
\aff { 10} {B} {BWWWWBWW} {B}
}\hskip 10pt
\vtop{\leftskip 0pt\parindent 0pt\hsize=\tabrulecol  %
\aff { 10} {B} {BWWWWBWW} {B}
\aff { 20} {B} {BBWWWBWW} {B}
\aff { 29} {B} {BWBWWBWW} {B}
\aff { 23} {B} {BWWBWBWW} {B}
\aff { 17} {B} {BWWWBBWW} {B}
}\hskip 10pt
\vtop{\leftskip 0pt\parindent 0pt\hsize=\tabrulecol  %
\aff { 44} {B} {BWWBBBWW} {B}
\aff { 48} {B} {BWBBWBWW} {B}
\aff { 52} {B} {BBBWWBWW} {B}
\aff { 20} {B} {BBWWWBWW} {B}
\aff { 10} {B} {BWWWWBWW} {B}
}\hskip 10pt
\vtop{\leftskip 0pt\parindent 0pt\hsize=\tabrulecol  %
\aff { 10} {B} {BWWWWBWW} {B}
\aff { 20} {B} {BBWWWBWW} {B}
\aff { 52} {B} {BBBWWBWW} {B}
\aff { 48} {B} {BWBBWBWW} {B}
\aff { 44} {B} {BWWBBBWW} {B}
}
\hfill} 
} 
\hfill} 
\vskip 9pt
\trfn
\vskip 5pt
\ligne{\hskip 5pt 2(7): \hfill   
\vtop{\leftskip 0pt\parindent 0pt\hsize=320pt  
\ligne{\hfill  
\vtop{\leftskip 0pt\parindent 0pt\hsize=\tabrulecol  %
\ligne{\hfill simple, $\curvearrowright$\hfill}
}\hskip 10pt
\vtop{\leftskip 0pt\parindent 0pt\hsize=\tabrulecol  %
\ligne{\hfill simple, $\curvearrowleft$\hfill}
}\hskip 10pt
\vtop{\leftskip 0pt\parindent 0pt\hsize=\tabrulecol  %
\ligne{\hfill double, $\curvearrowright$\hfill}
}\hskip 10pt
\vtop{\leftskip 0pt\parindent 0pt\hsize=\tabrulecol  %
\ligne{\hfill double, $\curvearrowleft$\hfill}
}
\hfill}  
\ligne{\hfill  
\vtop{\leftskip 0pt\parindent 0pt\hsize=\tabrulecol  %
\aff {  2} {B} {WWWWWWWW} {B}
\aff { 14} {B} {BWWWWWWW} {B}
\aff { 26} {B} {WBWWWWWW} {B}
\aff {  2} {B} {WWWWWWWW} {B}
}\hskip 10pt
\vtop{\leftskip 0pt\parindent 0pt\hsize=\tabrulecol  %
\aff {  2} {B} {WWWWWWWW} {B}
\aff { 26} {B} {WBWWWWWW} {B}
\aff { 14} {B} {BWWWWWWW} {B}
\aff {  2} {B} {WWWWWWWW} {B}
}\hskip 10pt
\vtop{\leftskip 0pt\parindent 0pt\hsize=\tabrulecol  %
\aff {  2} {B} {WWWWWWWW} {B}
\aff { 14} {B} {BWWWWWWW} {B}
\aff { 47} {B} {BBWWWWWW} {B}
\aff { 26} {B} {WBWWWWWW} {B}
\aff {  2} {B} {WWWWWWWW} {B}
}\hskip 10pt
\vtop{\leftskip 0pt\parindent 0pt\hsize=\tabrulecol  %
\aff {  2} {B} {WWWWWWWW} {B}
\aff { 26} {B} {WBWWWWWW} {B}
\aff { 47} {B} {BBWWWWWW} {B}
\aff { 14} {B} {BWWWWWWW} {B}
\aff {  2} {B} {WWWWWWWW} {B}
}
\hfill}  
}
\hfill} 
\vskip 9pt
\trfn
\vskip 8pt
}
\hfill}

It can be noted that all rules given in Table~\ref{rmiles} do not change the current
state of the cell. As they witness the passage of the locomotive, we cannot call them
conservative rules.

The difference of direction of the motion makes it that, both for the simple and the
double locomotive, that the same rules are used for both directions but in reverse
order. Also, in the case of the double locomotive, new rules appear when both cells
of the locomotive can be seen. At last, note for the cell~0(0), that the four 
neighbours~1(7), 1(6), 1(5) and~1(4) as represented in the rule constitute a window
over consecutive cells of the tracks occupying neighbours~2, 1, 8 and~7 in this order, 
see rules~31, 29, 32 and 34 respectively. Similar observations hold for the cell~1(8) 
with also a window of four consecutive cells of the tracks and for the cell~2(7) 
too, whose window consists of two cells only: cells~1(7) and 1(6), {\it i.e.}
neighbours~1 and 2 respectively.

\subsection{The rules for the fixed switch}\label{srfx}

We now turn to the study of the fixed switch, a passive structure as noted in 
Sub-section~\ref{struct}. Table~\ref{rfixed} gives new rules which are used for the 
crossing of the structure together with already used rules.

\ligne{\hfill 
\vtop{\leftskip 0pt\parindent 0pt\hsize=\tabruleli  
\begin{tab}\label{rfixed}
\leurre
Rules for the crossing of a fixed switch.
\end{tab}
\vskip-2pt
\trfn
\vskip 8pt
\ligne{\hfill simple locomotive\hfill}
\ligne{\hfill   
\vtop{\leftskip 0pt\parindent 0pt\hsize=\tabrulecol  
\aff { 71} {W} {BBWBWBBW} {W} 
\aff { 72} {B} {WBWBWWWW} {B} 
\aff { 73} {W} {WWWWBWBB} {W} 
\aff { 74} {W} {BWBWWWWW} {W} 
\aff { 75} {B} {WWWBBWWW} {B} 
}\hskip 10pt
\vtop{\leftskip 0pt\parindent 0pt\hsize=\tabrulecol  
\aff { 76} {W} {BBBWWWWB} {W} 
\aff { 77} {B} {WWWBWBWW} {B} 
\aff { 78} {W} {WBBWWWWB} {W} 
\aff { 79} {B} {WWWBWWBW} {B} 
\raff { 80} {W} {BBWBBBBW} {B} 
}\hskip 10pt
\vtop{\leftskip 0pt\parindent 0pt\hsize=\tabrulecol  
\aff { 81} {B} {WBWWWWWB} {B} 
\aff { 82} {B} {WWWBWWWB} {B} 
\raff { 83} {B} {BBWBWBBW} {W} 
\aff { 84} {W} {BBWBWBBB} {W} 
\aff { 85} {B} {WWWBWWWW} {B} 
}\hskip 10pt
\vtop{\leftskip 0pt\parindent 0pt\hsize=\tabrulecol  
\aff { 86} {B} {WBWWWWBW} {B} 
\aff { 87} {B} {WBWWWBWW} {B} 
\aff { 88} {B} {WBWWBWWW} {B} 
\raff { 89} {W} {BBBBWBBW} {B} 
}
\hfill} 
\vskip 7pt 
\ligne{\hfill double locomotive\hfill}
\ligne{\hfill   
\vtop{\leftskip 0pt\parindent 0pt\hsize=\tabrulecol  
\aff { 90} {B} {WWWBBBWW} {B} 
\aff { 91} {B} {WWWBWBBW} {B} 
\aff { 92} {B} {WWWBWWBB} {B} 
}\hskip 10pt
\vtop{\leftskip 0pt\parindent 0pt\hsize=\tabrulecol  
\aff { 93} {B} {BBWBBBBW} {B} 
\aff { 94} {B} {BBWWWWWB} {B} 
\raff { 95} {B} {BBWBWBBB} {W} 
}\hskip 10pt
\vtop{\leftskip 0pt\parindent 0pt\hsize=\tabrulecol  
\aff { 96} {B} {WBWWWWBB} {B} 
\aff { 97} {B} {WBWWWBBW} {B} 
\aff { 98} {B} {WBWWBBWW} {B} 
}\hskip 10pt
\vtop{\leftskip 0pt\parindent 0pt\hsize=\tabrulecol  
\aff { 99} {B} {BBBBWBBW} {B} 
}
\hfill} 
\vskip 9pt
\trfn
\vskip 8pt
}
\hfill}

\ligne{\hfill
\vtop{\hsize=220pt
\begin{tab}\label{efxsg}\leurre
Execution of the rules for the fixed switch when a simple locomotive comes from
the left-hand side.
\end{tab}
\vskip-2pt
\trep
\vskip 8pt
\ligne{\hfill\HH{}
\HH{{4$_8$} }\HH{{5$_8$} }\HH{{2$_1$} }\HH{{1$_1$} }\HH{{0$_0$} }\HH{{1$_4$} }\HH{{2$_4$} }\HH{{5$_3$} }\HH{{4$_3$} }
\hfill}
\ligne{\hfill\HH{1}
\HH{41}\HH{\Rr{42}}\HH{\Rr{40}}\HH{12}\HH{71}\HH{16}\HH{16}\HH{16}\HH{12}\hfill}
\ligne{\hfill\HH{2}
\HH{11}\HH{22}\HH{\Rr{42}}\HH{\Rr{40}}\HH{71}\HH{16}\HH{16}\HH{16}\HH{12}\hfill}
\ligne{\hfill\HH{3}
\HH{11}\HH{12}\HH{22}\HH{\Rr{42}}\HH{\Rr{80}}\HH{16}\HH{16}\HH{16}\HH{12}\hfill}
\ligne{\hfill\HH{4}
\HH{11}\HH{12}\HH{12}\HH{22}\HH{\Rr{83}}\HH{\Rr{18}}\HH{16}\HH{16}\HH{12}\hfill}
\ligne{\hfill\HH{5}
\HH{11}\HH{12}\HH{12}\HH{12}\HH{84}\HH{\Rr{24}}\HH{\Rr{18}}\HH{16}\HH{12}\hfill}
\ligne{\hfill\HH{6}
\HH{11}\HH{12}\HH{12}\HH{12}\HH{71}\HH{30}\HH{\Rr{24}}\HH{\Rr{18}}\HH{12}\hfill}
\ligne{\hfill\HH{7}
\HH{11}\HH{12}\HH{12}\HH{12}\HH{71}\HH{16}\HH{30}\HH{\Rr{24}}\HH{\Rr{40}}\hfill}
\vskip 9pt
\trfn
\vskip 8pt
}
\hfill}

Table~\ref{efxsg} shows us that the rules we have seen for a simple locomotive on
the tracks are also used here: both the sequence 16, \Rr{18}, \Rr{24} and~30 and 
12, \Rr{40}, \Rr{42} and~22. For~0(0), the central cell of the switch, we have the 
sequence of rules: 71, \Rr{80}, \Rr{83}, 84:
\vskip 5pt
\ligne{\hfill
\haff { 71} {W} {BBWBWBBW} {W} \hskip 10pt
\hraff { 80} {W} {BBWBBBBW} {B} \hskip 10pt
\hraff { 83} {B} {BBWBWBBW} {W} \hskip 10pt
\haff { 84} {W} {BBWBWBBB} {W}
\hfill}
\vskip 5pt
We can see that rule~71 is the conservative rule of~0(0). The cell has five milestones
in its neighbours~1, 2, 4, 6 and~7. A locomotive arriving from the left is seen in
neighbour~5 as shown by rule~\Rr{80}. Rule~\Rr{83} witnesses that the simple locomotive is
in the cell: it has the same neighbourhood as rule~71. At last, rule~84, which can see the
locomotive in its neighbour~1, witnesses that the locomotive left the cell~0(0).

Table~\ref{efxsd} shows us that when the locomotive comes from the right, the sequence
71, \Rr{80}, \Rr{83}, 84 is replaced by 71, \Rr{89}, \Rr{83}, 84: rule~\Rr{80} is replaced 
by rule~{\Rr{89}} \laff {W} {BBBBWBBW} {B} {,} where we can see that the locomotive
is in neighbour~3, which corresponds to an arriving locomotive from the right.

\ligne{\hfill
\vtop{\hsize=220pt
\begin{tab}\label{efxsd}\leurre
Execution of the rules for the fixed switch when a simple locomotive comes from
the right-hand side.
\end{tab}
\vskip-2pt
\trep
\vskip 8pt
\ligne{\hfill\HH{}
\HH{{4$_6$} }\HH{{5$_6$} }\HH{{2$_7$} }\HH{{1$_7$} }\HH{{0$_0$} }\HH{{1$_4$} }\HH{{2$_4$} }\HH{{5$_3$} }\HH{{4$_3$} }
\hfill}
\ligne{\hfill\HH{1}
\HH{30}\HH{\Rr{42}}\HH{\Rr{40}}\HH{12}\HH{71}\HH{16}\HH{16}\HH{16}\HH{12}\hfill}
\ligne{\hfill\HH{2}
\HH{16}\HH{22}\HH{\Rr{42}}\HH{\Rr{40}}\HH{71}\HH{16}\HH{16}\HH{16}\HH{12}\hfill}
\ligne{\hfill\HH{3}
\HH{16}\HH{12}\HH{22}\HH{\Rr{42}}\HH{\Rr{89}}\HH{16}\HH{16}\HH{16}\HH{12}\hfill}
\ligne{\hfill\HH{4}
\HH{16}\HH{12}\HH{12}\HH{22}\HH{\Rr{83}}\HH{\Rr{18}}\HH{16}\HH{16}\HH{12}\hfill}
\ligne{\hfill\HH{5}
\HH{16}\HH{12}\HH{12}\HH{12}\HH{84}\HH{\Rr{24}}\HH{\Rr{18}}\HH{16}\HH{12}\hfill}
\ligne{\hfill\HH{6}
\HH{16}\HH{12}\HH{12}\HH{12}\HH{71}\HH{30}\HH{\Rr{24}}\HH{\Rr{18}}\HH{12}\hfill}
\ligne{\hfill\HH{7}
\HH{16}\HH{12}\HH{12}\HH{12}\HH{71}\HH{16}\HH{30}\HH{\Rr{24}}\HH{\Rr{40}}\hfill}
\vskip 9pt
\trfn
\vskip 8pt
}
\hfill}

\ligne{\hfill
\vtop{\hsize=220pt
\begin{tab}\label{efxd}\leurre
Execution of the rules for the fixed switch when a double locomotive crosses
the switch.
\end{tab}
\vskip-2pt
\trep
\vskip 8pt
\ligne{\hfill from the left-hand side\hfill}
\ligne{\hfill\HH{}
\HH{{4$_8$} }\HH{{5$_8$} }\HH{{2$_1$} }\HH{{1$_1$} }\HH{{0$_0$} }\HH{{1$_4$} }\HH{{2$_4$} }\HH{{5$_3$} }\HH{{4$_3$} }
\hfill}
\ligne{\hfill\HH{1}
\HH{41}\HH{\Rr{60}}\HH{59}\HH{\Rr{40}}\HH{71}\HH{16}\HH{16}\HH{16}\HH{12}\hfill}
\ligne{\hfill\HH{2}
\HH{11}\HH{22}\HH{\Rr{60}}\HH{59}\HH{\Rr{80}}\HH{16}\HH{16}\HH{16}\HH{12}\hfill}
\ligne{\hfill\HH{3}
\HH{11}\HH{12}\HH{22}\HH{\Rr{60}}\HH{93}\HH{\Rr{18}}\HH{16}\HH{16}\HH{12}\hfill}
\ligne{\hfill\HH{4}
\HH{11}\HH{12}\HH{12}\HH{22}\HH{\Rr{95}}\HH{45}\HH{\Rr{18}}\HH{16}\HH{12}\hfill}
\ligne{\hfill\HH{5}
\HH{11}\HH{12}\HH{12}\HH{12}\HH{84}\HH{\Rr{49}}\HH{45}\HH{\Rr{18}}\HH{12}\hfill}
\ligne{\hfill\HH{6}
\HH{11}\HH{12}\HH{12}\HH{12}\HH{71}\HH{30}\HH{\Rr{49}}\HH{45}\HH{\Rr{40}}\hfill}
\vskip 5pt
\trfn
\vskip 5pt
\ligne{\hfill from the right-hand side\hfill}
\ligne{\hfill\HH{}
\HH{{4$_6$} }\HH{{5$_6$} }\HH{{2$_7$} }\HH{{1$_7$} }\HH{{0$_0$} }\HH{{1$_4$} }\HH{{2$_4$} }\HH{{5$_3$} }\HH{{4$_3$} }
\hfill}
\ligne{\hfill\HH{1}
\HH{30}\HH{\Rr{60}}\HH{59}\HH{\Rr{40}}\HH{71}\HH{16}\HH{16}\HH{16}\HH{12}\hfill}
\ligne{\hfill\HH{2}
\HH{16}\HH{22}\HH{\Rr{60}}\HH{59}\HH{\Rr{89}}\HH{16}\HH{16}\HH{16}\HH{12}\hfill}
\ligne{\hfill\HH{3}
\HH{16}\HH{12}\HH{22}\HH{\Rr{60}}\HH{99}\HH{\Rr{18}}\HH{16}\HH{16}\HH{12}\hfill}
\ligne{\hfill\HH{4}
\HH{16}\HH{12}\HH{12}\HH{22}\HH{\Rr{95}}\HH{45}\HH{\Rr{18}}\HH{16}\HH{12}\hfill}
\ligne{\hfill\HH{5}
\HH{16}\HH{12}\HH{12}\HH{12}\HH{84}\HH{\Rr{49}}\HH{45}\HH{\Rr{18}}\HH{12}\hfill}
\ligne{\hfill\HH{6}
\HH{16}\HH{12}\HH{12}\HH{12}\HH{71}\HH{30}\HH{\Rr{49}}\HH{45}\HH{\Rr{40}}\hfill}
\vskip 9pt
\trfn
\vskip 8pt
}
\hfill}

In Table~\ref{efxd} we can see the application of the rules when a double locomotive
crosses the switch. The sequences of rules we have noticed for the three-milestoned
cells of the track are replaced by the corresponding ones for a double locomotive. 
Also, the sequence of cells for the cell~0(0) is a bit changed as follows:
\vskip 5pt
\ligne{\hfill
\haff { 93} {B} {BBWBBBBW} {B} \hskip 10pt
\haff { 99} {B} {BBBBWBBW} {B} \hskip 30pt
\hraff { 95} {B} {BBWBWBBB} {W}
\hfill}
\vskip 5pt
When the locomotive comes from the left, rule~\Rr{80} is replaced by rule~93, when it 
comes from the right, rule~\Rr{89} is replaced by rule~99. In both cases, rule~\Rr{83} 
is replaced by rule~\Rr{95} and then, rule~84 applies as in the case of a simple 
locomotive.

\vtop{
\ligne{\hfill
\includegraphics[scale=0.55]{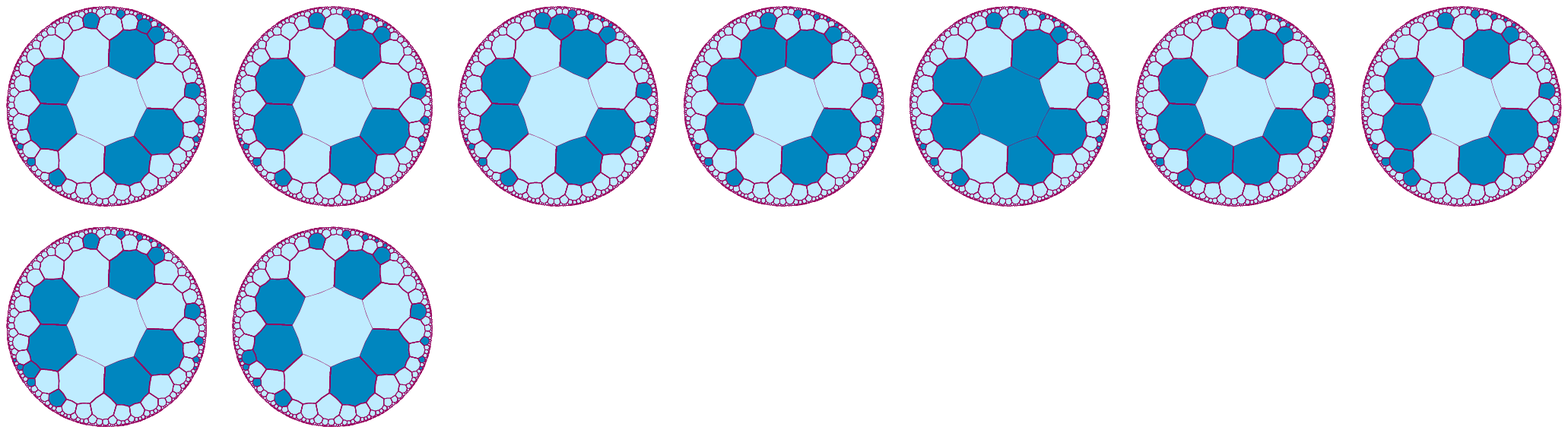}
\hfill}
\vspace{-15pt}
\ligne{\hfill
\includegraphics[scale=0.55]{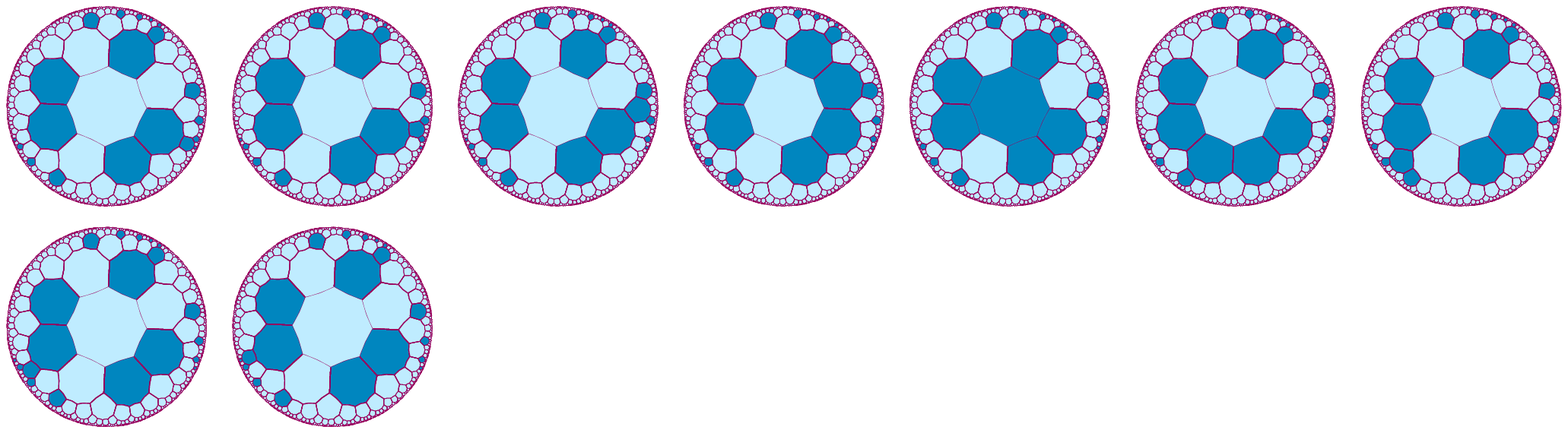}
\hfill}
\vspace{-15pt}
\ligne{\hfill
\includegraphics[scale=0.55]{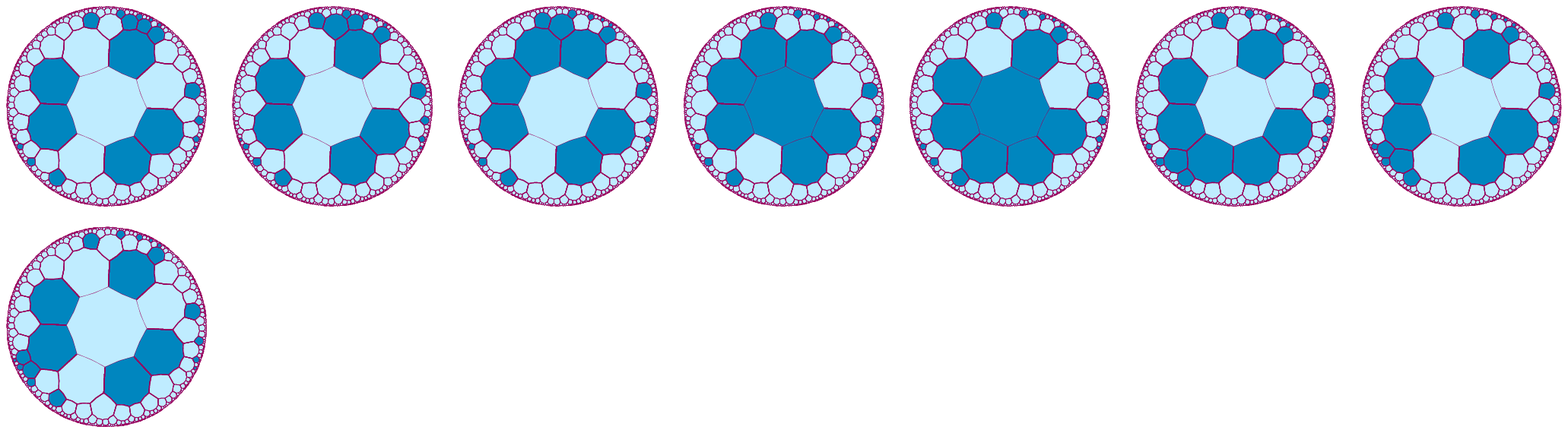}
\hfill}
\vspace{-15pt}
\ligne{\hfill
\includegraphics[scale=0.55]{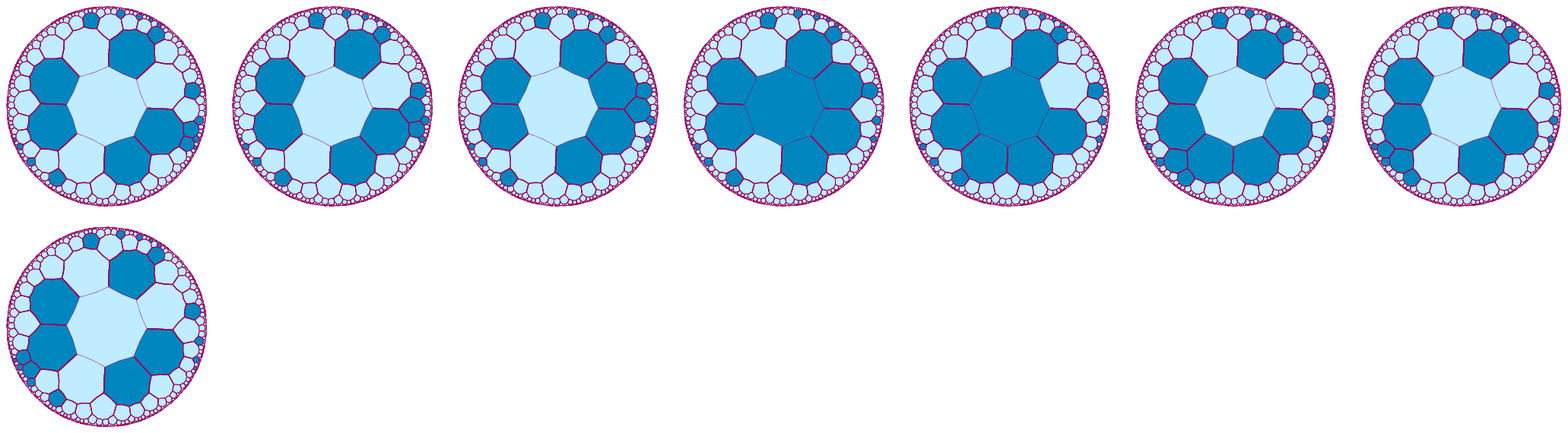}
\hfill}
\ligne{\hfill
\vtop{\leftskip 0pt\parindent 0pt\hsize=300pt
\begin{fig}\label{ffx}
\leurre
Illustration of the motion of a double locomotive in a clockwise motion, above,
and in a counterclockwise one, below.
\end{fig}
}
\hfill}
}

Figure~\ref{ffx} illustrates the four motions we have to consider for the
fixed switch for which Tables~\ref{efxsg}, \ref{efxsd} and~\ref{efxd} give the rules
used for such motions.

As in Sub-subsection~\ref{srtrack}, we conclude the sub-section with a study of a specific
milestone of the configuration: the cell~1(8), see Table~\ref{rmilfx}. Rule~85 is
the conservative rule of the cell. The cell itself is black, as required for a milestone.
It has a single black neighbour which is a milestone for the tracks arriving to the switch
from the left.

\ligne{\hfill  
\vtop{  
\begin{tab}\label{rmilfx}
\leurre
Rules witnessing the passage of the locomotive. 
Symbols~$\rightarrow$,
$\leftarrow$ indicate  a motion from the left, the right respectively.
\end{tab}
\vskip-2pt
\trep
\vskip 8pt
\ligne{\hskip 5pt 1(8): \hfill   
\vtop{\leftskip 0pt\parindent 0pt\hsize=320pt  
\ligne{\hfill  
\vtop{\leftskip 0pt\parindent 0pt\hsize=\tabrulecol  %
\ligne{\hfill simple, $\rightarrow$\hfill}
}\hskip 10pt
\vtop{\leftskip 0pt\parindent 0pt\hsize=\tabrulecol  %
\ligne{\hfill simple, $\leftarrow$\hfill}
}\hskip 10pt
\vtop{\leftskip 0pt\parindent 0pt\hsize=\tabrulecol  %
\ligne{\hfill double, $\rightarrow$\hfill}
}\hskip 10pt
\vtop{\leftskip 0pt\parindent 0pt\hsize=\tabrulecol  %
\ligne{\hfill double, $\leftarrow$\hfill}
}
\hfill}  
\ligne{\hfill 
\vtop{\leftskip 0pt\parindent 0pt\hsize=\tabrulecol  
\aff { 75} {B} {WWWBBWWW} {B}
\aff { 77} {B} {WWWBWBWW} {B}
\aff { 79} {B} {WWWBWWBW} {B}
\aff { 82} {B} {WWWBWWWB} {B}
\aff {  9} {B} {BWWBWWWW} {B}
\aff { 85} {B} {WWWBWWWW} {B}
}\hskip 10pt
\vtop{\leftskip 0pt\parindent 0pt\hsize=\tabrulecol  
\aff { 85} {B} {WWWBWWWW} {B}
\aff { 72} {B} {WBWBWWWW} {B}
\aff {  9} {B} {BWWBWWWW} {B}
\aff { 85} {B} {WWWBWWWW} {B}
}\hskip 10pt
\vtop{\leftskip 0pt\parindent 0pt\hsize=\tabrulecol  
\aff { 90} {B} {WWWBBBWW} {B}
\aff { 91} {B} {WWWBWBBW} {B}
\aff { 92} {B} {WWWBWWBB} {B}
\aff { 35} {B} {BWWBWWWB} {B}
\aff {  9} {B} {BWWBWWWW} {B}
\aff { 85} {B} {WWWBWWWW} {B}
}\hskip 10pt
\vtop{\leftskip 0pt\parindent 0pt\hsize=\tabrulecol  
\aff { 85} {B} {WWWBWWWW} {B}
\aff { 72} {B} {WBWBWWWW} {B}
\aff { 65} {B} {BBWBWWWW} {B}
\aff {  9} {B} {BWWBWWWW} {B}
\aff { 85} {B} {WWWBWWWW} {B}
}
\hfill} 
} 
\hfill} 
\vskip 9pt
\trfn
\vskip 8pt
}
\hfill}

The cell has two windows on the tracks: its neighbour~2 look at a cell of the 
tracks arriving to the switch from the right, see rules~72 in the case of a simple
locomotive. Neighbours~5, 6, 7 and~8 
offer a view on the tracks arriving to the switch from the left: see rules~75, 77, 79 
and~82 in the case of a simple locomotive. At last, neighbour~1 is
cell~0(0), the central cell of the switch. Note that rule~9 is common to all motions:
the last cell of the locomotive is seen in the cell~0(0) which is neighbour~1 for~1(8).

When the locomotive is double, the rules are different when the locomotive arrives
from the left: the window consists of four consecutive cells so that the two cells
are visible for three tops of the clock as witnessed by rules~90, 91 and 93. When the 
last cell of the locomotive is in 1(1), {\it i.e.} neighbour~8 for 1(8), the front
of the locomotive is in~0(0), which is neighbour~1 for~1(8), see rule~35. Note that
rule~35 appears in Table~\ref{rvoies}. That rule is also used by the cell~1(3) when
the locomotive is in the cell~1(4): it occurs in the motions on the track we already
studied. It also occurs here, when a simple locomotive crosses the switch whatever the
side from which it came. 

\subsection{The rules for the doubler and for the fork}\label{srdblfrk}

Presently, we look at the new rules involved by the doubler and by the fork. The rules for
the fork are mainly contained in those for the doubler as the doubler contains a fork.
It also contains a fixed switch so that the rules of Table~\ref{rfixed} should also be 
involved. The function of the fork is performed by the configuration of the cell~1(1)
together with the sequence of rules~100, \Rr{104}, and \Rr{106}:

\vskip 5pt
\ligne{\hfill
\haff {100} {W} {WBWBWBWB} {W} \hskip 10pt
\hraff {104} {W} {WBWBBBWB} {B} \hskip 10pt
\hraff {106} {B} {WBWBWBWB} {W} \hskip 10pt
\haff { 84} {W} {BBWBWBBB} {W} 
\hfill}
\vskip 5pt
Rule~100 is the conservative rule for~1(1). Note that two tracks have their origin
in~1(1). We chose side~1 to be shared with the cell~1(8). This is why rule~\Rr{104}
can see the arriving locomotive in neighbour~5. Rule~\Rr{106} witnesses the locomotive 
in the cell and makes it leave the cell. Rule~84 can see the locomotive in both 
neighbours~1, {\it i.e.} 1(8), and~7 which is~1(2). The left-hand side part of 
Table~\ref{edbl} shows us that starting from the application of rule~84, each locomotive 
goes on its way in the tracks: the locomotive created in~1(2) counterclockwise goes on, 
that created in~1(8) clockwise goes on.

\ligne{\hfill 
\vtop{\leftskip 0pt\parindent 0pt\hsize=\tabruleli  
\begin{tab}\label{rdblfork}
\leurre
Rules for the motion of the locomotive through the doubler and through the
fork.
\end{tab}
\vskip-2pt
\trep
\vskip 8pt
\ligne{\hfill   
\vtop{\leftskip 0pt\parindent 0pt\hsize=\tabrulecol  
\aff {100} {W} {WBWBWBWB} {W} 
\aff {101} {B} {WWWWWBWW} {B} 
\aff {102} {B} {WWBWWWWW} {B} 
\aff {103} {W} {WWWWWBWB} {W} 
\raff {104} {W} {WBWBBBWB} {B} 
\aff {105} {B} {WWWWBWBB} {B} 
}\hskip 10pt
\vtop{\leftskip 0pt\parindent 0pt\hsize=\tabrulecol  
\raff {106} {B} {WBWBWBWB} {W} 
\aff {107} {B} {WWWBWBBB} {B} 
\aff {108} {B} {WWBWWWBB} {B} 
\aff {109} {B} {BWBWWWWB} {B} 
\aff {110} {B} {BWWWWWBB} {B} 
\aff {111} {B} {BWWWBWWB} {B} 
}\hskip 10pt
\vtop{\leftskip 0pt\parindent 0pt\hsize=\tabrulecol  
\aff {112} {B} {BBWWWWBW} {B} 
\aff {113} {B} {BWBWWWWW} {B} 
\aff {114} {B} {BBWBBWWW} {B} 
\aff {115} {B} {WBBWWWWW} {B} 
\aff {116} {B} {BBWWBWWW} {B} 
\raff {117} {W} {WWWWWBBB} {B} 
}\hskip 10pt
\vtop{\leftskip 0pt\parindent 0pt\hsize=\tabrulecol  
\raff {118} {B} {WWWWWBBB} {W} 
\vskip 5pt
\ligne{\hfill fork\hfill}
\aff {119} {B} {WWWWBWWW} {B} 
\aff {120} {B} {WWBWWWBW} {B} 
}
\hfill} 
\vskip 9pt
\trfn
\vskip 8pt
}
\hfill}

\vskip 10pt

\ligne{\hfill
\vtop{\hsize=320pt
\begin{tab}\label{edbl}\leurre
Execution of the rules for the doubler. To left, around the cell~$1(1)$, the fork;
to right, around the cell~$4(4)$, the fixed switch.
\end{tab}
\vskip-2pt
\trep
\vskip 8pt
\ligne{\hfill
\vtop{\hsize=150pt
\ligne{\hfill\HH{}
\HH{{5$_1$} }\HH{{1$_1$} }\HH{{1$_2$} }\HH{{2$_3$} }\HH{{1$_8$} }\HH{{1$_7$} }
\hfill}
\ligne{\hfill\HH{1}
\HH{\Rr{24}}\HH{\Rr{104}}\HH{11}\HH{12}\HH{16}\HH{16}
\hfill}
\ligne{\hfill\HH{2}
\HH{30}\HH{\Rr{106}}\HH{\Rr{43}}\HH{12}\HH{\Rr{18}}\HH{16}
\hfill}
\ligne{\hfill\HH{3}
\HH{16}\HH{84}\HH{\Rr{39}}\HH{\Rr{40}}\HH{\Rr{24}}\HH{\Rr{18}}
\hfill}
\ligne{\hfill\HH{4}
\HH{16}\HH{100}\HH{41}\HH{\Rr{42}}\HH{30}\HH{\Rr{24}}
\hfill}
\ligne{\hfill\HH{5}
\HH{16}\HH{100}\HH{11}\HH{22}\HH{16}\HH{30}
\hfill}
}
\vtop{\hsize=150pt
\ligne{\hfill\HH{}
\HH{{2$_4$} }\HH{{3$_4$} }\HH{{4$_4$} }\HH{{5$_4$} }\HH{{2$_5$} }\HH{{15$_4$}}\hfill}
\ligne{\hfill\HH{5}
\HH{11}\HH{12}\HH{71}\HH{16}\HH{16}\HH{16}\hfill}
\ligne{\hfill\HH{6}
\HH{11}\HH{12}\HH{71}\HH{16}\HH{\Rr{18}}\HH{16}\hfill}
\ligne{\hfill\HH{7}
\HH{\Rr{43}}\HH{12}\HH{71}\HH{\Rr{18}}\HH{\Rr{24}}\HH{16}\hfill}
\ligne{\hfill\HH{8}
\HH{\Rr{39}}\HH{\Rr{40}}\HH{\Rr{89}}\HH{\Rr{24}}\HH{30}\HH{16}\hfill}
\ligne{\hfill\HH{9}
\HH{41}\HH{\Rr{60}}\HH{93}\HH{30}\HH{16}\HH{\Rr{18}}\hfill}
\ligne{\hfill\HH{10}
\HH{11}\HH{22}\HH{\Rr{95}}\HH{30}\HH{16}\HH{45}\hfill}
\ligne{\hfill\HH{11}
\HH{11}\HH{12}\HH{84}\HH{16}\HH{16}\HH{\Rr{49}}\hfill}
}
\hfill}
\vskip 9pt
\trfn
\vskip 8pt
}
\hfill}

\vtop{
\ligne{\hfill
\includegraphics[scale=0.55]{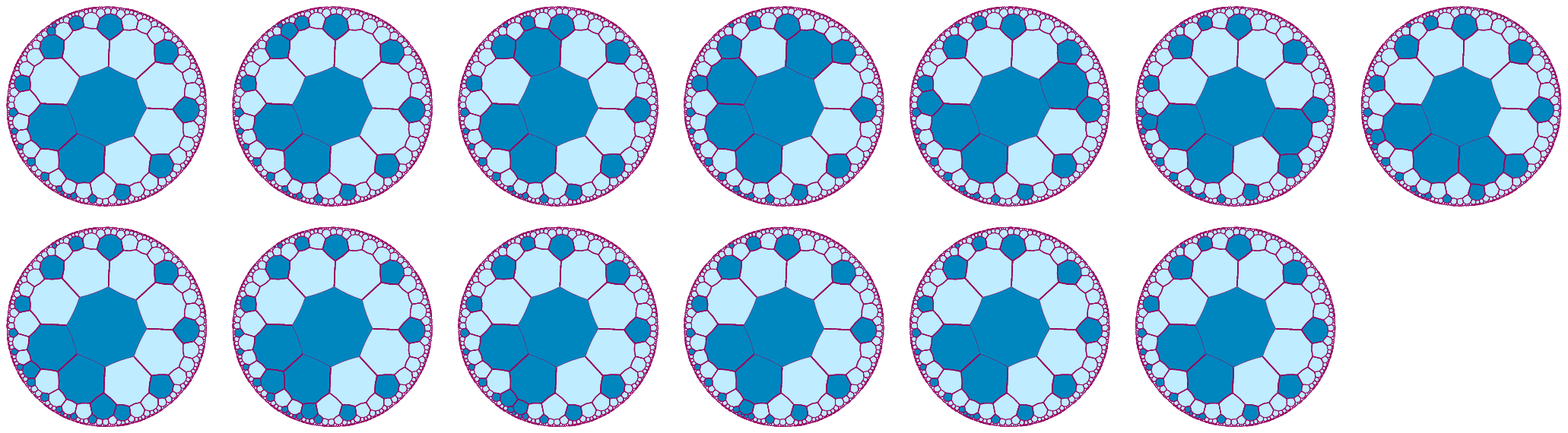}
\hfill}
\vspace{-15pt}
\ligne{\hfill
\includegraphics[scale=0.55]{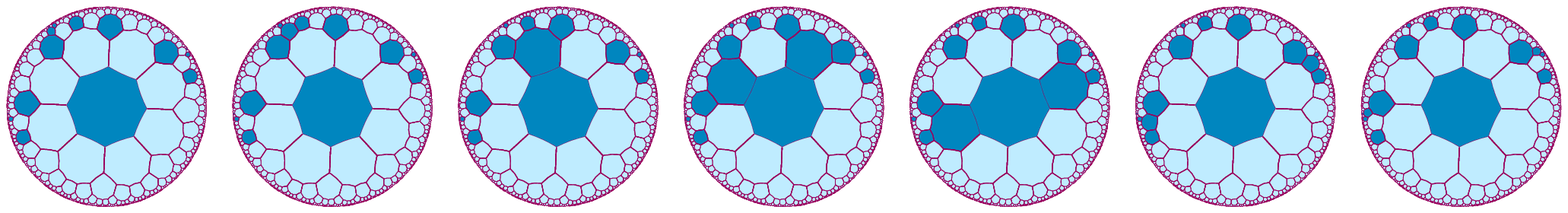}
\hfill}
\begin{fig}\label{fdblfrk}
Illustration of the crossing of the doubler, above, and of the fork, below, by the
locomotive.
\end{fig}
}
\ifnum 1=0 {
\vtop{
\ligne{\hfill\HH{}
\HH{{6$_2$} }\HH{{5$_1$} }\HH{{1$_1$} }\HH{{2$_1$} }\HH{{1$_3$} }
\HH{{3$_3$} }\HH{{9$_3$} }\HH{{1$_8$} }\HH{{1$_7$} }\HH{{5$_7$} }\HH{{6$_8$} }\hfill}
\ligne{\hfill\HH{1}
\HH{22}\HH{\Rr{24}}\HH{\Rr{104}}\HH{12}\HH{16}
\HH{16}\HH{12}\HH{16}\HH{12}\HH{12}\HH{16}\hfill}
\ligne{\hfill\HH{2}
\HH{12}\HH{30}\HH{\Rr{106}}\HH{\Rr{40}}\HH{16}
\HH{16}\HH{12}\HH{\Rr{18}}\HH{12}\HH{12}\HH{16}\hfill}
\ligne{\hfill\HH{3}
\HH{12}\HH{16}\HH{84}\HH{\Rr{42}}\HH{\Rr{18}}
\HH{16}\HH{12}\HH{\Rr{24}}\HH{\Rr{40}}\HH{12}\HH{16}\hfill}
\ligne{\hfill\HH{4}
\HH{12}\HH{16}\HH{100}\HH{22}\HH{\Rr{24}}
\HH{\Rr{18}}\HH{12}\HH{30}\HH{\Rr{42}}\HH{\Rr{40}}\HH{16}\hfill}
\ligne{\hfill\HH{5}
\HH{12}\HH{16}\HH{100}\HH{12}\HH{30}
\HH{\Rr{24}}\HH{\Rr{40}}\HH{16}\HH{22}\HH{\Rr{42}}\HH{\Rr{18}}\hfill}
}
} \fi

Note that rule~84 is also applied when a locomotive crosses a fixed switch: it witnesses
that the locomotive left the central cell of the switch. We also note that rule in
the right-hand side part of Table~\ref{edbl} which deals with the fixed-switch part of
the doubler.

The upper two rows of Figure~\ref{fdblfrk} illustrate the crossing of the doubler by a 
locomotive. The last row of the figure illustrates the working of the fork: we can see 
as it is the same as the one which is installed in the doubler. 

The doubler also contains a fixed switch, as already mentioned in Section~\ref{scenar}.
The right-hand side part of Table~\ref{edbl} shows us that the two locomotives
arrive at the cell~4(4) with a time difference of one top of the clock as already
mentioned. We can see that the locomotive which moved in a clockwise way arrive
sooner than the other one, so that at time~8, see Table~\ref{edbl}, we have
a double locomotive in the fixed switch with its front in the cell~4(4) and its 
rear in the cell~3(4), the neighbour~5 of the cell~4(4):
\vskip 5pt
\ligne{\hfill
\haff { 71} {W} {BBWBWBBW} {W} \hskip 2.5pt
\hraff { 89} {W} {BBBBWBBW} {B} \hskip 2.5pt
\haff { 93} {B} {BBWBBBBW} {B}  \hskip 2.5pt
\hraff { 95} {B} {BBWBWBBB} {W} \hskip 2.5pt
\haff { 84} {W} {BBWBWBBB} {W}
\hfill}
\vskip 5pt
Note that the sequence of rules is a bit different from that of Table~\ref{efxd}: the
situation is a kind of mix between the two parts of Table~\ref{efxd} as here, the
double locomotive arrives as two simple locomotive coming from \textit{both} sides of the
switch. 

We conclude this subsection by two sequences of rules concerning milestones: the cells
1(3) and~1(4). The former witnesses the passage of the locomotive around it on six steps.
The latter is a neighbour of~4(4), the central cell of the fixed switch, which is
also a neighbour of~3(4) and~5(4), cells which belong to the tracks arriving to~4(4).

\ligne{\hfill
\vtop{\hsize=\tabruleli
\begin{tab}\label{rmildbl}\leurre
Rules used at the milestones $1(3)$ and $1(4)$ of the doubler.
\end{tab}
\vskip-2pt
\trep
\vskip 8pt
\ligne{\hfill\hbox to 130pt{\hfill 1(3)\hfill}
\hbox to 130pt{\hfill 1(4)\hfill}\hfill}
\ligne{\hfill
\vtop{\leftskip 0pt\parindent 0pt\hsize=\tabrulecol  
\aff { 51} {B} {BWWWWWWB} {B}
\aff { 94} {B} {BBWWWWWB} {B}
\aff {109} {B} {BWBWWWWB} {B}
\aff { 35} {B} {BWWBWWWB} {B}
}
\vtop{\leftskip 0pt\parindent 0pt\hsize=\tabrulecol  
\aff {111} {B} {BWWWBWWB} {B}
\aff { 38} {B} {BWWWWBWB} {B}
\aff {110} {B} {BWWWWWBB} {B}
\aff { 51} {B} {BWWWWWWB} {B}
}
\hskip 20pt
\vtop{\leftskip 0pt\parindent 0pt\hsize=\tabrulecol  
\aff { 47} {B} {BBWWWWWW} {B}
\aff { 94} {B} {BBWWWWWB} {B}
\aff {112} {B} {BBWWWWBW} {B}
}
\vtop{\leftskip 0pt\parindent 0pt\hsize=\tabrulecol  
\aff { 52} {B} {BBBWWBWW} {B}
\aff {114} {B} {BBWBBWWW} {B}
\aff {116} {B} {BBWWBWWW} {B}
\aff { 47} {B} {BBWWWWWW} {B}
}
\hfill}
\vskip 9pt
\trfn
\vskip 8pt
}
\hfill}

Table~\ref{rmildbl} shows us the rules applied to the cells~1(3) and 1(4). The rules
for each cell are displayed on two columns and they follow each other from top to bottom
first in the left-hand side column and then in the right-hand side one. We can see 
that for~1(3), the window of six consecutive cells is clear as well as the motion
of the locomotive in the window. The cell 1(4) also has a window of six cells but their
display is a bit different and the rules do not concern the rightmost cell of the
window. In the first column devoted to~1(4), we can see the motion of the locomotive
which arrives from the right. The motion of both locomotives clearly appears in the 
second column. We can see the constitution of the double locomotive, see rule~52
followed by rule~114. We also can see
that the front corresponds to the right-hand side locomotive which arrives to~4(4)
as the first one. Rule~116 shows us that the rear of the double locomotive is
in~4(4). Table~\ref{edbl} shows us that when this is the case, rule~\Rr{49} shows us
that the front of the double locomotive arrived in~15(4). Table~\ref{rmildbl}
shows us that the cell~4(4) did the job more clearly than Figure~\ref{fdblfrk}.

\subsection{The rules for the selector}\label{srsel}

The selector is the last structure we need to implement round-abouts. The new rules needed
by the structure are given in Table~\ref{rsel} while the execution of the rules used
by the crossing of a locomotive are given in Table~\ref{esel}: the left-, right-hand side 
sub-table gives the rules used by a simple, double locomotive respectively.

In both sub-tables of Table~\ref{esel}, we can see that the track leading the locomotive
to the selector make use of motion rules examined in Sub-section~\ref{srtrack}.
We can see that the entrance to the selector, the cell 1(6), makes use of the same 
neighbourhood as the fixed switch. But as side~1 for 1(6) is different, the rules
managing the cell are different from those used in the fixed switch for the cell~0(0).

\ligne{\hfill 
\vtop{\leftskip 0pt\parindent 0pt\hsize=\tabruleli  
\begin{tab}\label{rsel}
\leurre
Rules for the locomotive through the selector.
\end{tab}
\vskip-2pt
\trep
\vskip 8pt
\ligne{\hfill simple locomotive\hfill}
\vskip 0pt
\ligne{\hfill   
\vtop{\leftskip 0pt\parindent 0pt\hsize=\tabrulecol  
\aff {121} {B} {WWWWBBBW} {B} 
\aff {122} {W} {WBWWBBWB} {W} 
\aff {123} {W} {WBBWBWBB} {W} 
\aff {124} {B} {WWBBBWWW} {B} 
\aff {125} {W} {WWWBWBWB} {W} 
\aff {126} {B} {WWWWBBBB} {B} 
\raff {127} {W} {WBBWBBBB} {B} 
}\hskip 10pt
\vtop{\leftskip 0pt\parindent 0pt\hsize=\tabrulecol  
\aff {128} {B} {WWWBBWBB} {B} 
\raff {129} {B} {WBBWBWBB} {W} 
\aff {130} {B} {BBBWWWWW} {B} 
\aff {131} {B} {WBBBBWWW} {B} 
\raff {132} {W} {WBBWWBWB} {B} 
\raff {133} {B} {BWWWBBBW} {W} 
\aff {134} {W} {BBBWBWBB} {W} 
}\hskip 10pt
\vtop{\leftskip 0pt\parindent 0pt\hsize=\tabrulecol  
\aff {135} {B} {BWBBBWWW} {B} 
\raff {136} {W} {WWWBBBWB} {B} 
\aff {137} {W} {BBWWBWWB} {W} 
\raff {138} {B} {WWWWWBWB} {W} 
\raff {139} {W} {WBWWBBBW} {B} 
\aff {140} {W} {WBWWWWBB} {W} 
\aff {141} {W} {WWBWBWBB} {W} 
}\hskip 10pt
\vtop{\leftskip 0pt\parindent 0pt\hsize=\tabrulecol  
\aff {142} {B} {WWBBBWWB} {B} 
\raff {143} {B} {WWWBWBWB} {W} 
\aff {144} {W} {WBWBWBBB} {W} 
\aff {145} {B} {WBWWBBBW} {B} 
\aff {146} {B} {WWBBBWBW} {B} 
\aff {147} {W} {WWWBWBBB} {W} 
}
\hfill} 
\vskip 9pt
\ligne{\hfill double locomotive\hfill}
\vskip 0pt
\ligne{\hfill 
\vtop{\leftskip 0pt\parindent 0pt\hsize=\tabrulecol  
\aff {148} {B} {BWBBWWWB} {B} 
\aff {149} {B} {WBBWBBBB} {B} 
\aff {150} {B} {BBBWWWWB} {B} 
\aff {151} {B} {BWWWBBBB} {B} 
}\hskip 10pt
\vtop{\leftskip 0pt\parindent 0pt\hsize=\tabrulecol  
\raff {152} {B} {BBBWBWBB} {W} 
\raff {153} {B} {BBBBBWWW} {W} 
\raff {154} {B} {BBWWBBWW} {W} 
\raff {155} {B} {WBBWWBWB} {W} 
}\hskip 10pt
\vtop{\leftskip 0pt\parindent 0pt\hsize=\tabrulecol  
\aff {156} {B} {BBWWBBBW} {B} 
\aff {157} {W} {BBBWBWBW} {W} 
\raff {158} {W} {BWBBBWWB} {B} 
}\hskip 10pt
\vtop{\leftskip 0pt\parindent 0pt\hsize=\tabrulecol  
\raff {159} {B} {WWWBBWWB} {W} 
\aff {160} {B} {WWBWBBBW} {B} 
}
\hfill} 
\vskip 9pt
\trfn
\vskip 8pt
} 
\hfill}

\ligne{\hfill
\vtop{\hsize=340pt
\begin{tab}\label{esel}\leurre
Execution of the rules for a locomotive passing through the selector.
\end{tab}
\vskip-2pt
\trep
\vskip 8pt
\ligne{\hfill
\vtop{\hsize=170pt
\ligne{\hfill simple locomotive\hfill}
\ligne{\hfill\HH{}
\HH{{6$_7$} }\HH{{5$_6$} }\HH{{1$_6$} }\HH{{0$_0$} }\HH{{1$_8$} }\HH{{2$_8$} }\HH{{9$_8$} }
\hfill}
\ligne{\hfill\HH{1}
\HH{22}\HH{\Rr{24}}\HH{\Rr{127}}\HH{16}\HH{125}\HH{12}\HH{16}
\hfill}
\ligne{\hfill\HH{2}
\HH{12}\HH{30}\HH{\Rr{129}}\HH{\Rr{18}}\HH{125}\HH{12}\HH{16}
\hfill}
\ligne{\hfill\HH{3}
\HH{12}\HH{16}\HH{134}\HH{\Rr{24}}\HH{\Rr{136}}\HH{12}\HH{16}
\hfill}
\ligne{\hfill\HH{4}
\HH{12}\HH{16}\HH{141}\HH{137}\HH{\Rr{143}}\HH{\Rr{40}}\HH{16}
\hfill}
\ligne{\hfill\HH{5}
\HH{12}\HH{16}\HH{123}\HH{16}\HH{147}\HH{\Rr{42}}\HH{\Rr{18}}
\hfill}
}
\vtop{\hsize=170pt
\ligne{\hfill double locomotive\hfill}
\ligne{\hfill\HH{}
\HH{{6$_7$} }\HH{{5$_6$} }\HH{{1$_6$} }\HH{{0$_0$} }\HH{{1$_4$} }\HH{{2$_5$} }\HH{{7$_5$} }
\hfill}
\ligne{\hfill\HH{1}
\HH{22}\HH{\Rr{49}}\HH{149}\HH{\Rr{18}}\HH{16}\HH{16}\HH{12}
\hfill}
\ligne{\hfill\HH{2}
\HH{12}\HH{30}\HH{\Rr{152}}\HH{45}\HH{\Rr{132}}\HH{16}\HH{12}
\hfill}
\ligne{\hfill\HH{3}
\HH{12}\HH{16}\HH{157}\HH{\Rr{154}}\HH{\Rr{155}}\HH{\Rr{18}}\HH{12}
\hfill}
\ligne{\hfill\HH{4}
\HH{12}\HH{16}\HH{123}\HH{16}\HH{30}\HH{\Rr{24}}\HH{\Rr{40}}
\hfill}
\ligne{\hfill\HH{5}
\HH{12}\HH{16}\HH{123}\HH{16}\HH{16}\HH{30}\HH{\Rr{42}}
\hfill}
}
\hfill}
\vskip 9pt
\trfn
\vskip 8pt
}
\hfill}

The sequence of rules used by~1(6) when a simple locomotive crosses it is given
by: 123, \Rr{127}, \Rr{129}, 134, 141:

\vskip 5pt
\ligne{\hfill
\haff {123} {W} {WBBWBWBB} {W} 
\hraff {127} {W} {WBBWBBBB} {B} 
\hraff {129} {B} {WBBWBWBB} {W} 
\haff {134} {W} {BBBWBWBB} {W} 
\haff {141} {W} {WWBWBWBB} {W} 
\hfill}
\vskip 5pt
Rule~123 is the conservative rule for the structure, see Figure~\ref{stab_sel}.
Rule~\Rr{127} can see the locomotive arrived in neighbour~6 which is the cell 5(6).
Rule~\Rr{129} can see the locomotive in the cell and it makes it leave the cell.
Rule~134 witnesses that the locomotive is now in the cell~0(0). Rule~141
can see that the cell 1(5) turned to white and then the next time, rule~123
again applies, witnessing that the cell~1(5) turned back to black. We shall see the 
effect of this flash issued by the cell~1(5).

The sequence of rules used by~1(6) when the locomotive is double is given
by the sequence: 123, \Rr{127}, 149, \Rr{152}, 157:
\vskip 5pt
\ligne{\hfill
\haff {123} {W} {WBBWBWBB} {W} 
\hraff {127} {W} {WBBWBBBB} {B} 
\haff {149} {B} {WBBWBBBB} {B} 
\hraff {152} {B} {BBBWBWBB} {W} 
\haff {157} {W} {BBBWBWBW} {W} 
\hfill}
\vskip 5pt
The first two rules are the same as previously. Now, after that, rule~149 is needed
to make the rear of the double locomotive enter the cell~1(6). Then, rule~\Rr{152}
makes 1(6) turn back to white as the locomotive, seen in neighbour~1, is now in the 
cell~0(0). Then, rule~157 can see the rear of the locomotive in~0(0) and
it also can see that the cell 1(7) turned to white as neighbour~8 in the rule is white.
As rule~123 again applies, there is no locomotive in~0(0) and 1(7) turned back
to black.

Table~\ref{rmilsel} gives the rules used by the cells~1(5) and~1(7) which we call the
\textbf{sensors} of the selector.

For each sensor, we give the rules for a passage of a simple locomotive
and those for the passage of a double one. Let us first look at the cell~1(5). The
conservative rule is rule~121 and the position of the three consecutive black neighbours
shows us that side~1 is shared with 0(0). Rule~126 witnesses that the locomotive
is in the cell~1(6). Rule~\Rr{133} can see the locomotive in 0(0) and as the locomotive
is simple, cell~1(6) is now white, so that the sensor 1(5) flashes: it becomes white.
Now, rule~\Rr{139} shows us that there is a locomotive in 1(4) and the rule makes the
cell 1(5) return to black. As cell~121 again applies, we can see that the locomotive
which was in 1(4) vanished as the cell 2(5) which belongs to the track starting from
1(4) is white. Also see Figure~\ref{fsel}.

\ligne{
\vtop{\leftskip 0pt\parindent 0pt\hsize=\tabruleli  
\begin{tab}\label{rmilsel}\leurre
The rules used by the sensors $1(5)$, to left, and $1(7)$, to right, of the
selector.
\end{tab}
\vskip-2pt
\trep
\vskip 8pt
\ligne{\hfill
\vtop{\leftskip 0pt\parindent 0pt\hsize=\tabrulecol  %
\ligne{\hfill simple\hfill}
\aff {121} {B} {WWWWBBBW} {B} 
\aff {126} {B} {WWWWBBBB} {B} 
\raff {133} {B} {BWWWBBBW} {W} 
\raff {139} {W} {WBWWBBBW} {B} 
\aff {121} {B} {WWWWBBBW} {B} 
}
\hskip 5pt
\vtop{\leftskip 0pt\parindent 0pt\hsize=\tabrulecol  
\ligne{\hfill double\hfill}
\aff {121} {B} {WWWWBBBW} {B} 
\aff {126} {B} {WWWWBBBB} {B} 
\aff {151} {B} {BWWWBBBB} {B}
\aff {156} {B} {BBWWBBBW} {B}
\aff {160} {B} {WWBWBBBW} {B}
\aff {121} {B} {WWWWBBBW} {B} 
}
\hskip 20pt
\vtop{\leftskip 0pt\parindent 0pt\hsize=\tabrulecol  
\ligne{\hfill simple\hfill}
\aff {124} {B} {WWBBBWWW} {B} 
\aff {131} {B} {WBBBBWWW} {B} 
\aff {135} {B} {BWBBBWWW} {B} 
\aff {142} {B} {WWBBBWWB} {B}
\aff {146} {B} {WWBBBWBW} {B}
\aff {124} {B} {WWBBBWWW} {B} 
}
\hskip 5pt
\vtop{\leftskip 0pt\parindent 0pt\hsize=\tabrulecol  
\ligne{\hfill double\hfill}
\aff {124} {B} {WWBBBWWW} {B} 
\aff {131} {B} {WBBBBWWW} {B} 
\raff {153} {B} {BBBBBWWW} {W}
\raff {158} {W} {BWBBBWWB} {B}
\aff {124} {B} {WWBBBWWW} {B} 
}
\hfill}
\vskip 9pt
\trfn
\vskip 8pt
}
\hfill
}

Consider the case when a double locomotive crosses the selector: for 1(5), it is the
second column of Table~\ref{rmilsel}. We can see that rule~151 detects that the
locomotive is double: both neighbours~1, 0(0), and~8, 1(6), are black. Next, rule~156
can see the rear of the locomotive in~0(0) and a new one in 1(4).
Then, rule~160 witnesses that there is no locomotive in 0(0), nor in 1(4), but
that the locomotive which was previously in~1(4) is now in~2(5). This means that when
the locomotive is simple, it does not go through the track starting from 1(4), but
when it is double, then a simple locomotive is sent on the track starting from~1(4).
This is illustrated by the first row of Figure~\ref{fsel}.

\vtop{
\ligne{\hfill
\includegraphics[scale=0.55]{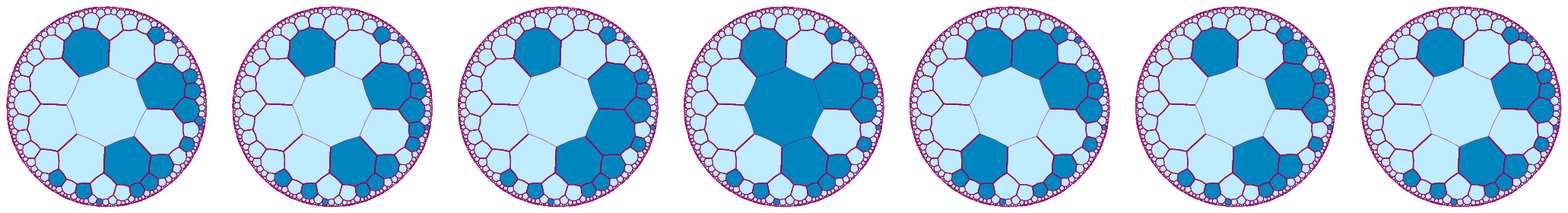}
\hfill}
\vspace{-15pt}
\ligne{\hfill
\includegraphics[scale=0.55]{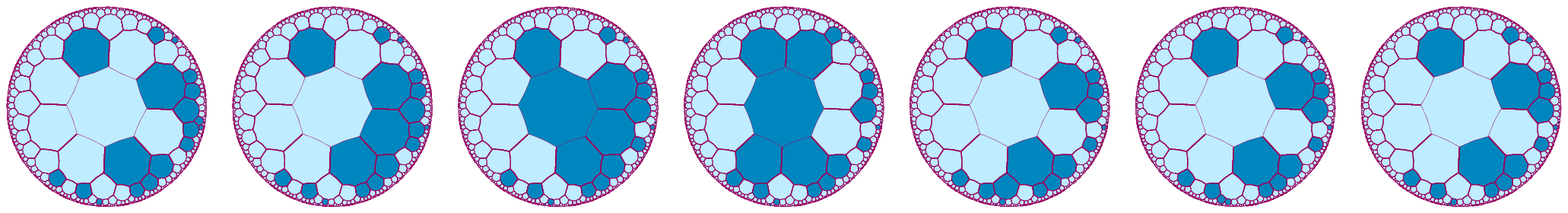}
\hfill}
\begin{fig}\label{fsel}\leurre
Illustration of the crossing of the selector: above, by a simple locomotive, 
below, by a double one.
\end{fig}
}

Let us now study what happens at the sensor~1(7). It has the same neighbourhood as the
sensor~1(5). Its side~1 is also shared with~0(0), but as the numbering is always defined
with the same orientation, rule~121 cannot apply and a new rule is needed: rule~124.
The detection of the front of the locomotive is detected by rule~131. Rule~135
detects that a simple locomotive arrived. Accordingly, the sensor remains black. After 
that, rules~142 and~146 witness that a simple locomotive was in 1(8) and 2(8) respectively:the locomotive goes on the track issued from 1(8). 

If a double locomotive arrives, rule~131 again applies and at the next step,
when the front of the locomotive is in~0(0), its rear is in~1(6)
which is detected by rule~\Rr{153} which makes the sensor flash: it becomes white.
Rule~\Rr{158} restores the black colour of the sensor and witnesses that a simple
locomotive is in 1(8) and that the rear of the double locomotive is in~0(0).
Rule~124 shows that, at the next time, there is no locomotive in~0(0), neither in~1(8)
nor in 2(8). This is illustrated by the second row of Figure~\ref{fsel}.
Accordingly, the rules up to the last one of Table~\ref{rsel} allow the selector
to work as indicated in Section~\ref{scenar}.

\subsection{The rules for the controller}\label{srctrl}

Let us now consider the rules for the controller of the active switches. The rules are
displayed by Table~\ref{rctrl}. As mentioned in the table itself, the two columns in 
the left-hand side deal with the passage of the locomotive while the last column deals
with the change of colour of the controller. We remind the reader that the colour
of the controller is the colour of the cell~1(3) in Figure~\ref{stab_ctrl}. 
Table~\ref{ectrlb} and Figure~\ref{rctrl} illustrate the crossing of a black controller 
by the locomotive.
All cells of the track obey the rules we have considered for the tracks for 
three-milestoned rules but the cell 1(4) for which specific rules are used, although
it is also a three-milestoned cell. Let us look at those specific rules as well as 
the rules used for the cell~1(3), also specific rules, all of them repeated in
Table~\ref{rctrl43}.

Let us first consider the cell 1(4) when the locomotive crosses a black controller.
Rule~164 is the conservative rule. The next cell on the track is 2(5) so that the
side shared by those cells is side~1 for~1(4), whence the new rules 164, \Rr{167}, 
\Rr{168} and 169 although 1(4) is a three-milestoned cell. The reason is that in the usual
situation of the tracks, the exit occurs through a cell which stands in between 
two milestones. This is not the case here: this requires specific rules. To these
rules, we could append corresponding rules for the case of a double locomotive:
\laff {B} {WBBBWBWW} {B} {} and \laff {B} {BBWBWBWW} {W} {.} 
Note that the following rules:
\vskip 5pt
\ligne{\hfill
\haff  {   } {W} {WWWBWBWB} {W} \hskip 10pt
\hraff {136} {W} {WWWBBBWB} {B} \hskip 10pt
\hraff {143} {B} {WWWBWBWB} {W} \hskip 10pt
\haff  {   } {W} {BWWBWBWB} {W}
\hfill
}
\vskip 5pt
\noindent
are compatible with the present rules. To these rules, it would also be possible
to use rules for a double locomotive, looking these cells as tracks only:
rules \laff {B} {WWWBBBWB} {B} {} and \laff {B} {BWWBWBWB} {B} {.}
As can be expected, cell~1(4) is not affected
when the locomotive arrives to a white controller. Rule~173 is the conservative rule
in that case and it is always applied during such a passage: the locomotive does not
even arrive to~0(0).

\ligne{\hfill
\vtop{\leftskip 0pt\parindent 0pt\hsize=300pt
\begin{tab}\label{rctrl43}\leurre
Rules for~$1(4)$ and for~$1(3)$, when the locomotive passes
and when the signal passes.
\end{tab}
\vskip-2pt
\trep
\vskip 8pt
\ligne{\hfill \hbox to 130pt{\hfill cell 1(4)\hfill}
\hfill\hbox to 130pt{\hfill cell 1(3)\hfill}\hfill}
\vskip 3pt
\trfn
\vskip 5pt
\ligne{\hfill
\vtop{\leftskip 0pt\parindent 0pt\hsize=\tabrulecol  
\ligne{\hfill locomotive\hfill}
\ligne{\hfill black\hfill}
\aff {164} {W} {WBWBWBWW} {W}
\raff {167} {W} {WBBBWBWW} {B}
\raff {168} {B} {WBWBWBWW} {W}
\aff {169} {W} {BBWBWBWW} {W}
\aff {164} {W} {WBWBWBWW} {W}
\vskip 5pt
\ligne{\hfill white\hfill}
\aff {173} {W} {WBWWWBWW} {W}
}
\hskip 5pt
\vtop{\leftskip 0pt\parindent 0pt\hsize=\tabrulecol  
\ligne{\hfill signal\hfill}
\ligne{\hfill black\hfill}
\aff {164} {W} {WBWBWBWW} {W}
\aff {180} {W} {WBWBBBWW} {W}
\aff {173} {W} {WBWWWBWW} {W}
\vskip 5pt
\ligne{\hfill white\hfill}
\aff {173} {W} {WBWWWBWW} {W}
\aff {178} {W} {WBWWBBWW} {W}
\aff {164} {W} {WBWBWBWW} {W}
}
\hskip 20pt
\vtop{\leftskip 0pt\parindent 0pt\hsize=\tabrulecol  
\ligne{\hfill locomotive\hfill}
\ligne{\hfill black\hfill}
\aff { 68} {B} {BBBBWWWW} {B}
\aff {163} {B} {BBBBWWBW} {B}
\aff {166} {B} {BBBBWBWW} {B}
\aff { 68} {B} {BBBBWWWW} {B}
\vskip 5pt
\ligne{\hfill white\hfill}
\aff {172} {W} {BBBBWWWW} {W}
}
\hskip 5pt
\vtop{\leftskip 0pt\parindent 0pt\hsize=\tabrulecol  
\ligne{\hfill signal\hfill}
\ligne{\hfill black\hfill}
\aff { 68} {B} {BBBBWWWW} {B}
\raff {153} {B} {BBBBBWWW} {W}
\aff {172} {W} {BBBBWWWW} {W}
\vskip 5pt
\ligne{\hfill white\hfill}
\aff {172} {W} {BBBBWWWW} {W}
\raff {177} {W} {BBBBBWWW} {B}
\aff { 68} {B} {BBBBWWWW} {B}
}
\hfill}
\vskip 9pt
\trfn
\vskip 8pt
}
\hfill}

\ligne{\hfill 
\vtop{\leftskip 0pt\parindent 0pt\hsize=\tabrulecol
\advance\hsize by\tabrulecol\advance \hsize by \tabrulecol
\advance\hsize by 40pt  
\begin{tab}\label{rctrl}
\leurre
Rules for the control: passage of the locomotive and signal for changing the selected
track.
\end{tab}
\vskip-2pt
\trep
\vskip 8pt
\ligne{\hfill passage of the locomotive\hfill
\hbox to \tabrulecol{\hfill signal\hfill}}
\vskip 0pt
\ligne{\hfill   
\vtop{\leftskip 0pt\parindent 0pt\hsize=\tabrulecol  
\ligne{\hfill black\hfill}
\aff {161} {W} {WWWWWWBB} {W}
\aff {162} {W} {BWWWWWBB} {W}
\aff {163} {B} {BBBBWWBW} {B}
\aff {164} {W} {WBWBWBWW} {W}
\aff {165} {W} {WWWBWWWW} {W}
\aff {166} {B} {BBBBWBWW} {B}
\raff {167} {W} {WBBBWBWW} {B}
\raff {168} {B} {WBWBWBWW} {W}
\aff {169} {W} {BBWBWBWW} {W}
}\hskip 10pt
\vtop{\leftskip 0pt\parindent 0pt\hsize=\tabrulecol  
\ligne{\hfill white\hfill}
\aff {170} {W} {WBWBWWWW} {W} 
\aff {171} {W} {WWWWWWBW} {W} 
\aff {172} {W} {BBBBWWWW} {W} 
\aff {173} {W} {WBWWWBWW} {W} 
\aff {174} {W} {WWBWBWWB} {W} 
}\hskip 10pt
\vtop{\leftskip 0pt\parindent 0pt\hsize=\tabrulecol  
\ligne{\hfill $W \rightarrow B$\hfill}
\aff {175} {W} {WBBBWWWW} {W}
\raff {176} {W} {WWBBBWWB} {B}
\raff {177} {W} {BBBBBWWW} {B}
\aff {178} {W} {WBWWBBWW} {W}
\raff {179} {B} {WWBWBWWB} {W}
\vskip 5pt
\ligne{\hfill $B \rightarrow W$\hfill}
\aff {180} {W} {WBWBBBWW} {W} 
}
\hfill}
\vskip 9pt
\trfn
\vskip 8pt
}
\hfill}

When the signal arrives to the controller in order to change the colour in the
cell~1(3), the cell~1(4) remains white and it simply witnesses the arrival of the
signal to the cell~2(4) and the corresponding change in 1(3): rule~180 when
1(3) is black, rule~178 when it is white. 

Presently, consider the case of the cell 1(3). Rule~68 is the conservative rule
when the controller is black, rule~172 is the conservative rule when it is white.
When 1(3) is black, rule~163 can see the locomotive in 0(0), its neighbour~7,
and rule~166 can see it in 1(4), its neighbour~6. When 1(3) is white, it remains 
unchanged when the locomotive arrives and remains under rule~172 for the same
reason as the cell 1(4) remains under rule~173.

The cell~1(3) can see the signal when it is in the cell 2(4) only: the cell 6(4) is
not visible from~1(3). The change from black to white is performed by rule~\Rr{153}
and the change from white to black is performed by rule~\Rr{177}. In both cases, the 
signal is seen in neighbour~5, see Table~\ref{rctrl43}. See Table~\ref{ectrlo} too. 

\ligne{\hfill
\vtop{\hsize=220pt
\begin{tab}\label{ectrlb}\leurre
Execution of the rules used during the traversal of a black controller by the locomotive.
\end{tab}
\vskip-2pt
\trep
\vskip 8pt
\ligne{\hfill\HH{}
\HH{{6$_7$} }\HH{{2$_7$} }\HH{{1$_6$} }\HH{{0$_0$} }\HH{{1$_4$} }\HH{{2$_5$} }\HH{{3$_5$} }\HH{{10$_5$}}\HH{{1$_3$} }\hfill}
\ligne{\hfill\HH{1}
\HH{22}\HH{\Rr{42}}\HH{\Rr{18}}\HH{12}\HH{164}\HH{16}\HH{16}\HH{16}\HH{68}
\hfill}
\ligne{\hfill\HH{2}
\HH{12}\HH{22}\HH{\Rr{24}}\HH{\Rr{40}}\HH{164}\HH{16}\HH{16}\HH{16}\HH{68}
\hfill}
\ligne{\hfill\HH{3}
\HH{12}\HH{12}\HH{30}\HH{\Rr{42}}\HH{\Rr{167}}\HH{16}\HH{16}\HH{16}\HH{163}
\hfill}
\ligne{\hfill\HH{4}
\HH{12}\HH{12}\HH{16}\HH{22}\HH{\Rr{168}}\HH{\Rr{132}}\HH{16}\HH{16}\HH{166}
\hfill}
\ligne{\hfill\HH{5}
\HH{12}\HH{12}\HH{16}\HH{12}\HH{169}\HH{\Rr{24}}\HH{\Rr{18}}\HH{16}\HH{68}
\hfill}
\vskip 9pt
\trfn
\vskip 8pt
}
\hfill}

\ligne{\hfill
\vtop{\leftskip 0pt\parindent 0pt\hsize=310pt
\begin{tab}\label{ectrlo}\leurre
Execution of the rules when the locomotive arrives to a white controller and
when the signal for changing the colour arrives.
\end{tab}
\vskip-2pt
\trep
\vskip 8pt
\ligne{\hfill
\hbox to 125pt{\hfill locomotive, 1(3) white\hfill}
\hskip 5pt \hbox to 180pt{\hfill signal for changing the colour\hfill}
\hfill}
\vskip 3pt
\ligne{\hfill
\vtop{\hsize=125pt
\ligne{\hfill\HH{}
\HH{{6$_7$} }\HH{{2$_7$} }\HH{{1$_6$} }\HH{{0$_0$} }\hfill}
\ligne{\hfill\HH{1}
\HH{22}\HH{\Rr{42}}\HH{\Rr{18}}\HH{170}
\hfill}
\ligne{\hfill\HH{2}
\HH{12}\HH{22}\HH{\Rr{24}}\HH{175}
\hfill}
\ligne{\hfill\HH{3}
\HH{12}\HH{12}\HH{16}\HH{170}
\hfill}
}
\hskip 5pt
\vtop{\hsize=85pt
\ligne{\hfill\tt W $\rightarrow$ B\hfill}
\vskip 3pt
\ligne{\hfill\HH{}
\HH{{1$_3$} }\HH{{6$_4$} }\HH{{2$_4$} }\hfill}
\ligne{\hfill\HH{1}
\HH{\Rr{177}}\HH{30}\HH{\Rr{179}}
\hfill}
\ligne{\hfill\HH{2}
\HH{68}\HH{16}\HH{11}
\hfill}
}
\hskip 5pt
\vtop{\hsize=85pt
\ligne{\hfill\tt B $\rightarrow$ W\hfill}
\ligne{\hfill\HH{}
\HH{{1$_3$} }\HH{{6$_4$} }\HH{{2$_4$} }\hfill}
\ligne{\hfill\HH{1}
\HH{\Rr{153}}\HH{30}\HH{\Rr{39}}
\hfill}
\ligne{\hfill\HH{2}
\HH{172}\HH{16}\HH{174}
\hfill}
}
\hfill}
\vskip 9pt
\trfn
\vskip 8pt
}
\hfill}

\vskip 30pt
\vtop{
\ligne{\hfill
\includegraphics[scale=0.55]{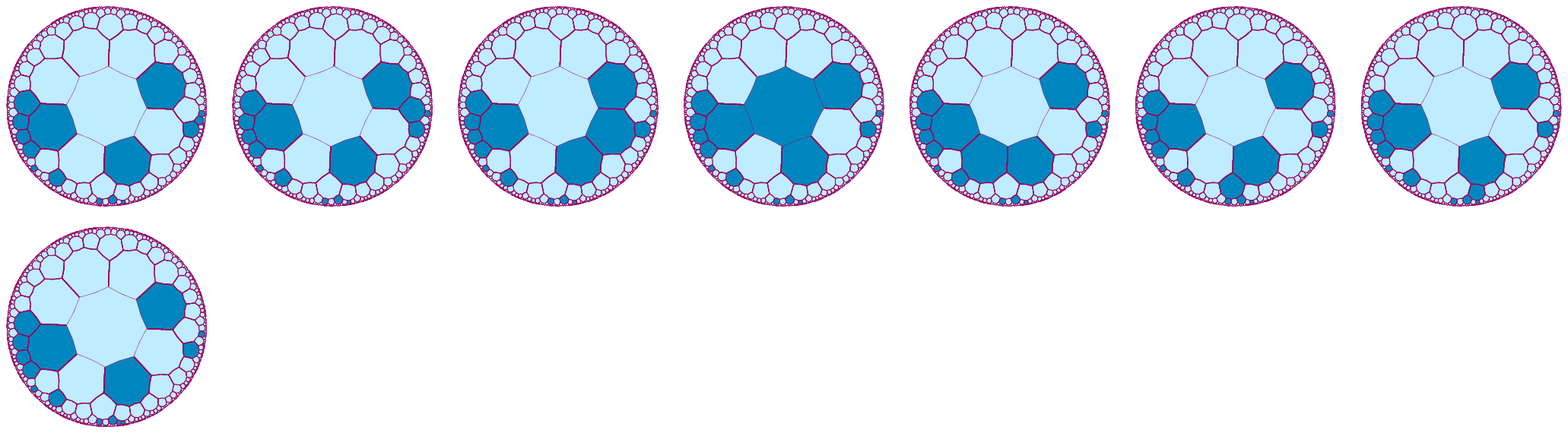}
\hfill}
\vspace{-70pt}
\ligne{\hskip 75pt
\includegraphics[scale=0.55]{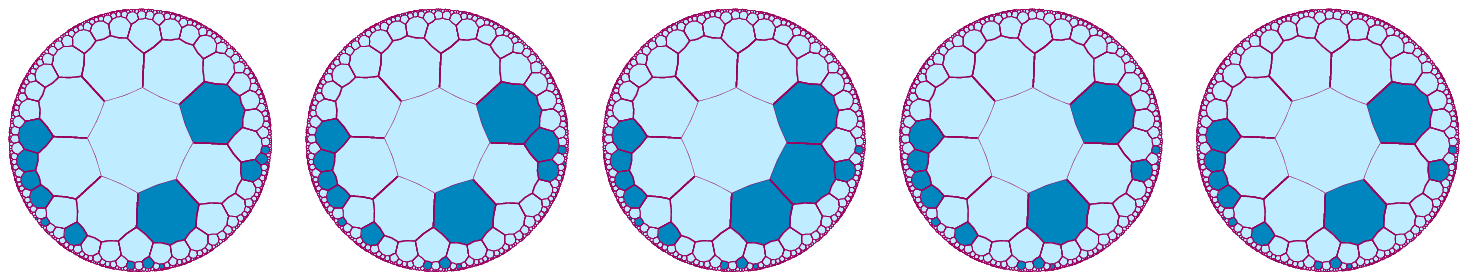}
\hfill}
\begin{fig}\label{fctrl}\leurre
Illustration of the crossing of the controller by the locomotive: above and first
figure of the second row, when it 
is black, below after the first figure, when it is white.
\end{fig}
}

Figure~\ref{fctrl} illustrates the crossing of controller by the locomotive
in both cases, according to the colour of~1(3). Figure~\ref{fctrls} illustrates the
arrival of the signal for changing the colour of the controller.

\vtop{
\ligne{\hskip 70pt
\includegraphics[scale=0.55]{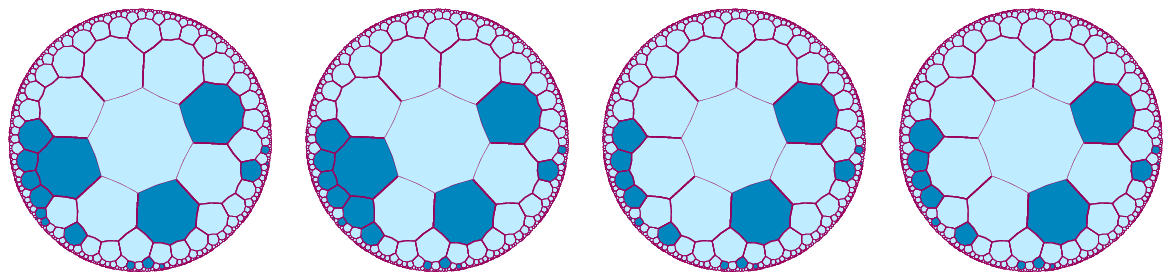}
\hfill}
\vspace{-20pt}
\ligne{\hskip 70pt
\includegraphics[scale=0.55]{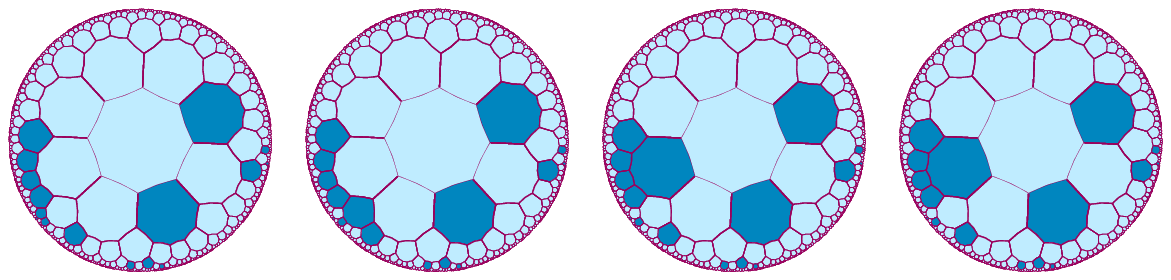}
\hfill}
\begin{fig}\label{fctrls}\leurre
Illustration of the arrival of the signal to the controller. Above: from black to white;
below: from white to black.
\end{fig}
}

\subsection{The rules for the sensor}\label{srcapt}

In this last subsection of Section~\ref{rules}, we examine the rules which manage the
working of the sensor, the specific control structure of the passive memory switch.
Sub-section~\ref{sctrlcapt} in Section~\ref{scenar} explained the working of
the structure, pointing at the differences between the controller and the sensor
illustrated by Figures~\ref{stab_ctrl} and~\ref{stab_capt}.

Table~\ref{rcapt} illustrates the rules which have to be appended to the already
examined ones in order to make the structure working as expected.

\ligne{\hfill
\vtop{\leftskip 0pt\parindent 0pt\hsize=\tabrulecol
\advance\hsize by\tabrulecol\advance \hsize by \tabrulecol
\advance\hsize by 40pt  
\begin{tab}\label{rcapt}\leurre
Rules for the sensor of the passive memory switch.
\end{tab}
\vskip-2pt
\trep
\vskip 8pt
\ligne{\hfill
\vtop{\leftskip 0pt\parindent 0pt\hsize=\tabrulecol  
\ligne{\hfill black\hfill}
\aff {181} {W} {WBBWBBWW} {W} 
\aff {182} {W} {WWBWWWWW} {W} 
\aff {183} {W} {WBBBWBWB} {W} 
}\hskip 10pt
\vtop{\leftskip 0pt\parindent 0pt\hsize=\tabrulecol  
\ligne{\hfill white\hfill}
\aff {184} {W} {WWBWBBWW} {W} 
\aff {185} {W} {WBWWWWBW} {W} 
\raff {186} {W} {BBBBWWBW} {B} 
\aff {187} {W} {BBWWWWBW} {W} 
\aff {188} {W} {BBWBWBWB} {W} 
}\hskip 10pt
\vtop{\leftskip 0pt\parindent 0pt\hsize=\tabrulecol  
\ligne{\hfill\tt B $\rightarrow$ W\hfill}
\raff {189} {W} {WBBBBBWW} {B} 
\raff {190} {B} {WBBWBBWW} {W} 
\aff {191} {W} {WBBWWWWW} {W} 
}
\hfill}
\vskip 9pt
\trfn
\vskip 8pt
}
\hfill}
\vskip 5pt

A few rules are appended to the 180 previous rules examined in the previous sections.
As can be seen in the comparison of Figures~\ref{stab_ctrl} and~\ref{stab_capt},
many rules used for the controller are also used for the sensor. As an example, as long
as the sensor is white, the rules executed in the cells of the tracks when the locomotive
passes are the same as those used in the same action when the controller is black,
see Tables~\ref{ectrlb} and~\ref{ecaptw}.

Besides the fact that the cell 1(3) is no more considered in Table~\ref{ecaptw}, 
the difference is in the rules concerning the cell~0(0) starting from time~4.
The reason is that the passage of the locomotive through the white sensor
transforms it into a black sensor, see the first row of Figure~\ref{fcapt}.

Let us look at what happens at the cell~0(0). Instead of the rules 12, \Rr{40}, \Rr{42}, 
22 for the passage of a simple locomotive, where rule~12 is the conservative rule
of 0(0) in the white sensor, we have the sequence 12, \Rr{40}, \Rr{42}, 188, 100:
\vskip 5pt
\ligne{\hfill \haff {188} {W} {BBWBWBWB} {W} \hskip 20pt \haff {100} {W} {WBWBWBWB} {W}
\hfill}
\vskip 5pt
\noindent
where rule~100 is the conservative rule for~0(0) in a black sensor. The comparison
with rule~12, \laff {W} {WBWBWWWB} {W} {,} shows us that the locomotive is in 1(4),
which means that it left the cell~0(0), and that the cell 1(1), neighbour~6 of~0(0),
is now black. Rule~100 confirms that the cell~1(1) remains black: the sensor turned to
black.

\ligne{\hfill
\vtop{\leftskip 0pt\parindent 0pt\hsize=200pt
\begin{tab}\label{ecaptw}\leurre
Execution of the rules when the sensor is white and then a locomotive passes.
\end{tab}
\vskip-2pt
\trep
\vskip 8pt
\ligne{\hfill\HH{}
\HH{{6$_7$} }\HH{{2$_7$} }\HH{{1$_6$} }\HH{{0$_0$} }\HH{{1$_4$} }\HH{{2$_5$} }\HH{{3$_5$} }\HH{{10$_5$}}\hfill}
\ligne{\hfill\HH{1}
\HH{22}\HH{\Rr{42}}\HH{\Rr{18}}\HH{12}\HH{164}\HH{16}\HH{16}\HH{16}
\hfill}
\ligne{\hfill\HH{2}
\HH{12}\HH{22}\HH{\Rr{24}}\HH{\Rr{40}}\HH{164}\HH{16}\HH{16}\HH{16}
\hfill}
\ligne{\hfill\HH{3}
\HH{12}\HH{12}\HH{30}\HH{\Rr{42}}\HH{\Rr{167}}\HH{16}\HH{16}\HH{16}
\hfill}
\ligne{\hfill\HH{4}
\HH{12}\HH{12}\HH{16}\HH{188}\HH{\Rr{168}}\HH{\Rr{132}}\HH{16}\HH{16}
\hfill}
\ligne{\hfill\HH{5}
\HH{12}\HH{12}\HH{16}\HH{100}\HH{169}\HH{\Rr{24}}\HH{\Rr{18}}\HH{16}
\hfill}
\vskip 9pt
\trfn
\vskip 8pt
}
\hfill}

\vtop{
\ligne{\hfill
\includegraphics[scale=0.55]{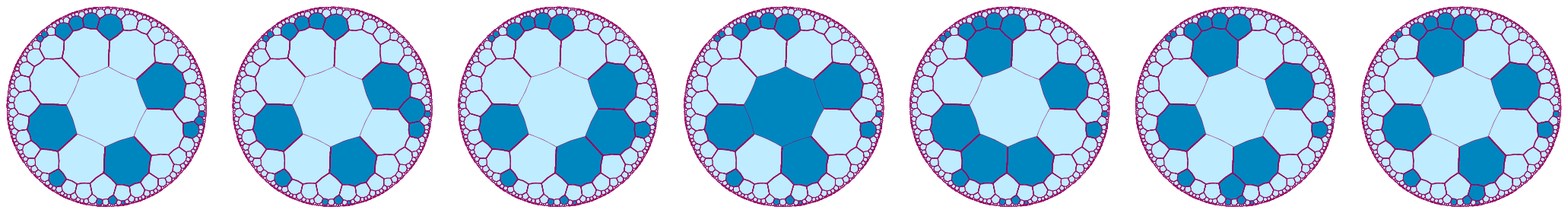}
\hfill}
\vspace{-20pt}
\ligne{\hfill
\includegraphics[scale=0.55]{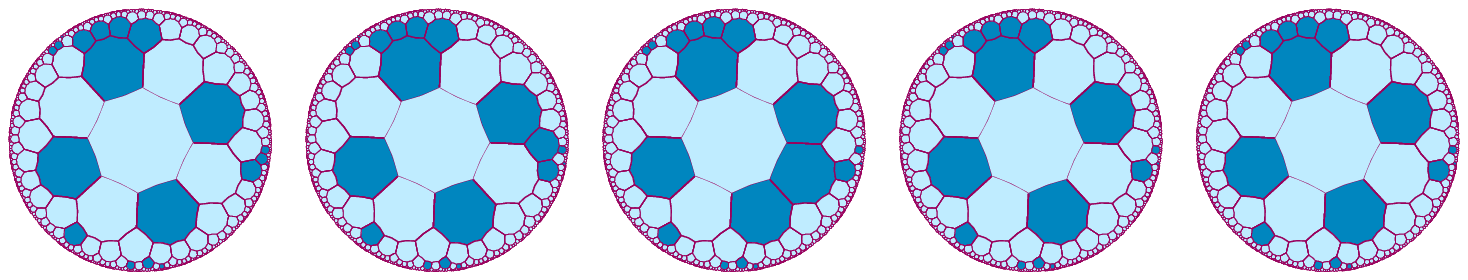}
\hfill}
\vspace{-20pt}
\ligne{\hfill
\includegraphics[scale=0.55]{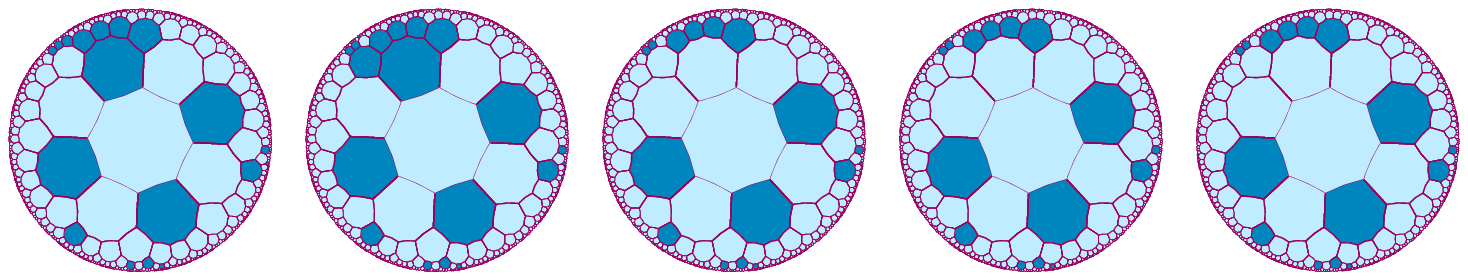}
\hfill}
\begin{fig}\label{fcapt}\leurre
Illustration of the working of the sensor. First row: passage of the locomotive;
second row: the locomotive is stopped; third row: the signal changes a black sensor
to a white one.
\end{fig}
}

Table~\ref{ecaptb} displays the rules in the cells of the tracks when the locomotive
arrives at a black sensor as well as the rules used by the cells of the track
which conveys the signal for changing the colour. 

Let us look at the cell~1(1) which gives its colour to the sensor. The conservative rule
for~1(1) is rule~68 when it is black, rule~172 when it is white. The change of colour are
induced by rule~\Rr{186} for \hbox{\tt W $\rightarrow$ B} and by rule~\Rr{153} for
\hbox{\tt B $\rightarrow$ W}. We have already seen rule~\Rr{153} which operates the same
change for the cell~1(3) which gives the colour of the controller. In
rule~\Rr{186}, \laff {W} {BBBBWWBW} {B} {,} we can see that the change is triggered
by the arrival of the locomotive at the cell~0(0) which is neighbour~7 in 1(1).
In the controller, 0(0) is also neighbour~7 for 1(3). But rule~\Rr{186} cannot be
applied in the controller: when 1(3) is white in the controller, the locomotive
never occurs in~0(0). In the controller, the change was triggered by the
occurrence of the signal in neighbour~5. This is why we again find rule~\Rr{153} here for 
turning the colour from black to white, as in the case of the controller. Note that
the conservative rules for the cell 1(1) are the same as those for the cell 1(3) in
the controller.

\ligne{\hfill
\vtop{\leftskip 0pt\parindent 0pt\hsize=250pt
\begin{tab}\label{ecaptb}\leurre
Execution of the rules for the black sensor, for the locomotive and for the signal.
\end{tab}
\vskip-2pt
\trep
\vskip 8pt
\ligne{\hfill
\vtop{\leftskip 0pt\parindent 0pt\hsize=130pt
\ligne{\hfill locomotive\hfill}
\ligne{\hfill\HH{}
\HH{{6$_7$} }\HH{{2$_7$} }\HH{{1$_6$} }\HH{{0$_0$} }\HH{{1$_4$} }\hfill}
\ligne{\hfill\HH{1}
\HH{22}\HH{\Rr{42}}\HH{\Rr{18}}\HH{100}\HH{164}
\hfill}
\ligne{\hfill\HH{2}
\HH{12}\HH{22}\HH{\Rr{24}}\HH{183}\HH{164}
\hfill}
\ligne{\hfill\HH{3}
\HH{12}\HH{12}\HH{16}\HH{100}\HH{164}
\hfill}
}
\hskip 10pt
\vtop{\leftskip 0pt\parindent 0pt\hsize=110pt
\ligne{\hfill signal\hfill}
\ligne{\hfill\HH{}
\HH{{10$_5$}}\HH{{1$_1$} }\HH{{6$_2$} }\HH{{2$_2$} }\hfill}
\ligne{\hfill\HH{1}
\HH{16}\HH{\Rr{153}}\HH{30}\HH{\Rr{190}} \hfill}
\ligne{\hfill\HH{2}
\HH{16}\HH{172}\HH{16}\HH{184} \hfill}
}
\hfill}
\vskip 9pt
\trfn
\vskip 8pt
}
\hfill}

We observe that the sub-table of Table~\ref{ecaptb} ruling the signal is almost the
same as the corresponding one in Table~\ref{ectrlo}. The difference is in the cell
at which the signal arrives: the cell 2(4) in the controller, the cell 2(2) here, in
the sensor. For the black sensor, the rules for 2(2) are 181, \Rr{189}, \Rr{190}, 184:
\vskip 5pt
\ligne{\hfill
\haff {181} {W} {WBBWBBWW} {W} \hskip 10pt
\hraff {189} {W} {WBBBBBWW} {B} \hskip 10pt
\hraff {190} {B} {WBBWBBWW} {W} \hskip 10pt
\haff {184} {W} {WWBWBBWW} {W}
\hfill}
\vskip 5pt
Rule~181 is the conservative rule for 2(2) in a black sensor. Rule~\Rr{189} can see the
arriving signal in neighbour~4 which is~6(2) and it makes it enter the cell. 
Rule~\Rr{190} restores the white colour of~2(2). But the content of 1(1), neighbour~2,
is changed so that the new conservative rule is rule~181, when the sensor is white.
Why is the cell 2(2) of the sensor different from the cell~2(4) of the controller?
The reason is that according to the natural definition of side~1 which would be the side
shared with~1(2), a new rule making the cell enter 2(2) could be accepted but then,
the rule \laff {B} {WBBBBWWW} {W} {} would be in conflict with 
rule~131, \laff {B} {WBBBBWWW} {B} {} which is not compatible as a motion rule.

Let us remark that Table~\ref{ecaptb} shows us that the locomotive is stopped by a
black sensor: the conservative rule for 0(0) in the black sensor is rule~100, as already
noticed. Also note that we already met the rule in the doubler and in the fork. At some
point rule~183, \laff {W} {WBBBWBWB} {W} {,} is used. The rule shows us that 0(0) can
see the arriving locomotive in 1(6), its neighbour~3, so that it remains white
as the sensor, neighbour~6, is black. This is
enough to cancel that locomotive as the usual motion rules of the tracks erase it
from the cell 2(7), see Figure~\ref{stab_capt}.

We completed the examination of the rules for the sensor.
Accordingly, Theorem~\ref{letheo} is proved.
\hfill\boxempty

\section*{Conclusion}

The result is close to the best: the tessellation $\{7,3\}$ is the tessellation 
$\{p,3\}$ where $p$ has the smallest value as possible for the hyperbolic plane. Is
it possible to implement the same model there? The question seems to be very difficult.

\end{document}